\documentclass[a4paper,fleqn]{cas-sc}

\usepackage{amssymb}
\usepackage{amsmath}
\usepackage{caption}
\usepackage{booktabs} 
\usepackage{longtable}
\usepackage{multirow} 
\usepackage{array} 
\usepackage{graphicx}
\usepackage{subcaption}
\usepackage{float}
\usepackage{hyperref}
\usepackage{threeparttable}
\usepackage{pifont}
 
\usepackage[ruled,vlined]{algorithm2e} 

\usepackage[numbers]{natbib}

\def\tsc#1{\csdef{#1}{\textsc{\lowercase{#1}}\xspace}}
\tsc{WGM}
\tsc{QE}

\newtheorem{definition}{Definition}
\newtheorem{theorem}{Theorem}
\newtheorem{lemma}[theorem]{Lemma}
\newdefinition{rmk}{Remark}
\newproof{pf}{Proof}
\newproof{pot}{Proof of Theorem \ref{thm}}

\begin{document}
\let\WriteBookmarks\relax
\def\floatpagepagefraction{1}
\def\textpagefraction{.001}

\shorttitle{Federated CCTR Prediction with LLM Augmentation}    

\shortauthors{Jiangcheng Qin et~al.}  

\title [mode = title]{Federated Cross-Domain Click-Through Rate Prediction With Large Language Model Augmentation}  



%

\author[1,2]{Jiangcheng Qin}[orcid=0000-0002-7705-6040]
\ead{qjc@zjhu.edu.cn}

\credit{Conceptualization, Data curation, Methodology, Software, Visualization, Validation, Formal analysis, Writing - original draft}

\affiliation[1]{organization={Ningbo University},
    addressline={No.818 Fenghua Road}, 
    city={Ningbo},
    postcode={315211}, 
    state={Zhejiang},
    country={China}}

\affiliation[2]{organization={Huzhou Key Laboratory of Sound Resource Data Mining and Intelligent Service Technology, Huzhou University},
    addressline={No.759 Erhuan East Road}, 
    city={Huzhou},
    postcode={313000}, 
    state={Zhejiang},
    country={China}}

\author[1]{Xueyuan Zhang}
\ead{1711082129@nbu.edu.cn}
\credit{Data curation, Methodology, Formal analysis}

\author[1]{Baisong Liu}[orcid=0000-0003-0401-6037]
\ead{lbs@nbu.edu.cn}
\credit{Funding acquisition, Supervision, Project administration, Writing - review \& editing}
\cormark[1]

\author[1]{Jiangbo Qian}[orcid=0000-0003-4245-3246]
\ead{qianjiangbo@nbu.edu.cn}
\credit{Funding acquisition, Supervision, Validation}

\author[1]{Yangyang Wang}
\ead{wangyangyang@nbu.edu.cn}
\credit{Writing - review \& editing, Formal analysis, Software}

\cortext[cor1]{Corresponding author}


\begin{abstract}
Accurately predicting click-through rates (CTR) under stringent privacy constraints poses profound challenges, particularly when user-item interactions are sparse and fragmented across domains. Conventional cross-domain CTR (CCTR) methods frequently assume homogeneous feature spaces and rely on centralized data sharing, neglecting complex inter-domain discrepancies and the subtle trade-offs imposed by privacy-preserving protocols. Here, we present Federated Cross-Domain CTR Prediction with Large Language Model Augmentation (FedCCTR-LM), a federated framework engineered to address these limitations by synchronizing data augmentation, representation disentanglement, and adaptive privacy protection. Our approach integrates three core innovations. First, the Privacy-Preserving Augmentation Network (PrivAugNet) employs large language models to enrich user and item representations and expand interaction sequences, mitigating data sparsity and feature incompleteness. Second, the Independent Domain-Specific Transformer with Contrastive Learning (IDST-CL) module disentangles domain-specific and shared user preferences, employing intra-domain representation alignment (IDRA) and cross-domain representation disentanglement (CDRD) to refine the learned embeddings and enhance knowledge transfer across domains. Finally, the Adaptive Local Differential Privacy (AdaLDP) mechanism dynamically calibrates noise injection to achieve an optimal balance between rigorous privacy guarantees and predictive accuracy. Empirical evaluations on four real-world datasets demonstrate that FedCCTR-LM substantially outperforms existing baselines, offering robust, privacy-preserving, and generalizable cross-domain CTR prediction in heterogeneous, federated environments.
\end{abstract}


\begin{highlights}
\item \textbf{A Paradigm Shift in Federated CCTR Prediction}: Our FedCCTR-LM is the first framework to seamlessly integrate Cross-domain CTR prediction within a federated environment, fostering privacy-conscious collaboration across domains. 
\item \textbf{Innovative Augmentation Network}: We propose PrivAugNet, leveraging LLMs to counter data sparsity and heterogeneity across domains, yielding richer and more consistent representations without centralizing raw data. 
\item \textbf{Advanced Representation Learning}: Our IDST-CL model disentangles domain-specific and cross-domain preferences through contrastive learning, enabling efficient knowledge transfer and addressing representational inconsistencies across domains. 
\item \textbf{Dynamic Privacy Mechanism}: Our AdaLDP mechanism dynamically balances privacy constraints and model utility, preserving sensitive user information while maintaining predictive fidelity.
\end{highlights}


\maketitle

\section{Introduction}\label{introduction}
\begin{figure*}[!t]
    \centering
    \includegraphics[trim={0cm 20cm 0cm 0cm}, clip, width=0.8\textwidth]{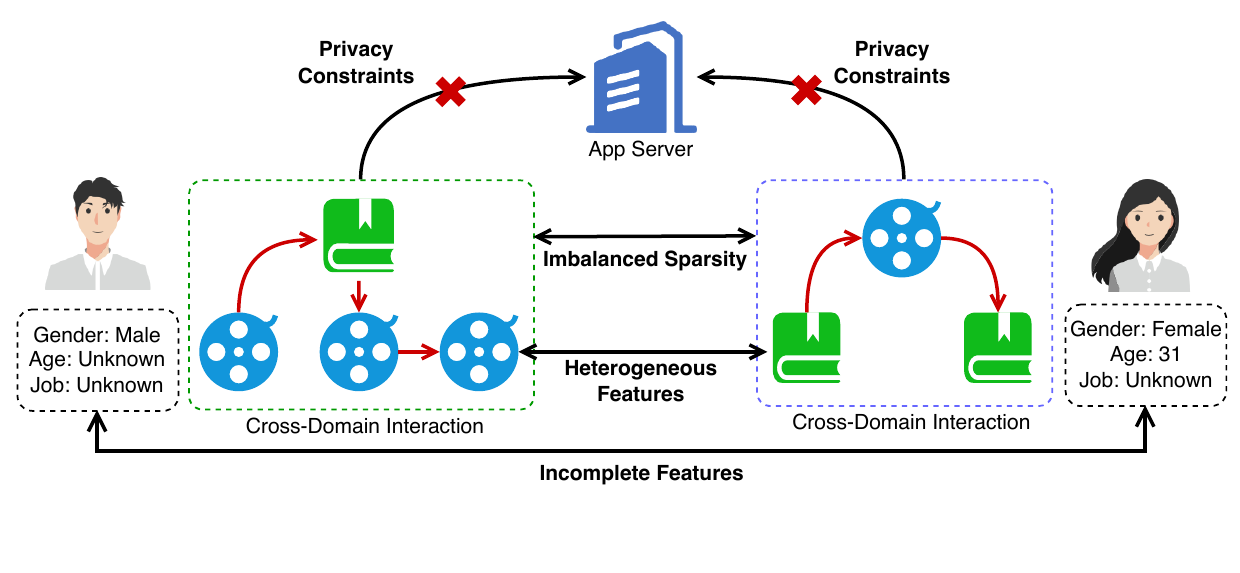}
    \caption{A Toy Illustration of Cross-Domain CTR Prediction Challenges: Non-I.I.D. Data Distributions, Knowledge Transfer Barriers, and Privacy-Utility Trade-offs.}
    \label{fig:toy_illustration}
\end{figure*}

Click-through rate (CTR) prediction serves as a cornerstone in modern recommender systems, guiding personalized content delivery and informing strategic decisions in e-commerce, streaming services, and online advertising\citep{guo2017deepfm, cheng2016wide, zhou2019deep, chen2019behavior}. By estimating the probability that a user will engage with a given item, CTR models enhance user satisfaction and maximize platform revenues. Yet, beneath their widespread adoption, these models face persistent and evolving challenges. Chief among these are data sparsity—where limited interactions obscure user preferences\citep{chen2024enhancing, Gao2022Causal, yin2020overcoming}-and feature heterogeneity—where structurally distinct domains and feature heterogeneity\citep{li2019heterogeneous,li2019heterogeneous2,li2018heterogeneous3}. The complexity intensifies when bridging \textit{cross-domain} environments. As users seamlessly traverse platforms—such as consuming both books and movies in an online marketplace\citep{lin2019cross} or using a streaming service alongside a shopping app\citep{gao2021cross}—their preferences fragment across disparate systems, each with its own data distribution and behavioral dynamics. Exploiting these multi-domain signals to refine CTR prediction holds considerable promise, yet conventional approaches often falter under privacy constraints and irregular data distributions.

Early forays into CCTR have leveraged matrix factorization\citep{sahu2018matrix,kuang2021deep,zhu2018deep,li2019heterogeneous}, transfer learning\citep{li2021dual,liu2023continual,zhang2023collaborative,li2020atlrec,li2019heterogeneous2}, graph-based models\citep{lei2021DA-GCN,li2023preference,ma2022mixed,guo2022time,ouyang2021mobile}, and sequential modeling approaches\citep{li2021dual,ma2024triple,cao2022c2dsr} to share knowledge between domains. While these methods can mitigate sparsity, they typically assume centralized data pools and uniform access to sensitive interaction logs. Such assumptions grow increasingly untenable under stringent data protection regulations—exemplified by GDPR\footnote{GDPR (General Data Protection Regulation): \url{https://gdpr-info.eu/}} and CCPA\footnote{CCPA (California Consumer Privacy Act): \url{https://oag.ca.gov/privacy/ccpa}}—and amid rising competition for proprietary user data. Beyond these legal and organizational constraints, existing models often fail to capture the intrinsic heterogeneity and nuanced feature distributions across domains, resulting in suboptimal, biased representations. These shortcomings underscore a pressing need for federated solutions that operate without centralizing sensitive data, and that can gracefully navigate the intricacies of inter-domain alignment, user-level heterogeneity, and strict privacy guarantees.

Federated learning (FL)\citep{mcmahan2017communication} has emerged as a promising paradigm to address these challenges. By decentralizing model training, FL ensures that sensitive user interaction data remains localized, effectively addressing privacy concerns and regulatory constraints while enabling collaborative knowledge sharing across domains\citep{zhang2024feddcsr,tian2024privacy}. However, despite these advantages, the application of FL to CCTR prediction is far from straightforward. CCTR emphasizes enhancing prediction performance within local domains by leveraging globally learned knowledge, but interaction data across domains is inherently heterogeneous\citep{tan2022towards,khan2017cross}, reflecting distinct characteristics such as genre preferences in books versus actor specifications in movies. These discrepancies, coupled with data sparsity and imbalanced distributions\citep{wang2021deconfounded}, complicate the construction of unified representations that can simultaneously capture domain-specific nuances and shared cross-domain insights. Additionally, the private nature of interaction data limits centralized access, necessitating solutions for distributed information fusion and adaptive knowledge transfer across domains. Privacy-preserving mechanisms, such as differential privacy\citep{abadi2016deep}, provide theoretical guarantees but often obscure critical patterns with excessive noise, while rigid privacy constraints may hinder generalization. Addressing these complexities requires balancing domain-specific personalization, robust cross-domain knowledge transfer, and stringent privacy preservation, highlighting critical research gaps in federated CCTR prediction.

This study focuses on addressing three key research challenges arising in federated CCTR prediction:

\begin{enumerate}
    \item \textbf{Non-I.I.D. Data Distributions:} \textit{How can federated CCTR prediction effectively address the challenges posed by non-I.I.D. data distributions, particularly for users with incomplete, heterogeneous, and imbalanced interactions sparsity across domains?}
    \item \textbf{Cross-Domain Knowledge Transfer:} \textit{How can federated CCTR model effectively facilitate the transfer of knowledge across domains while preserving domain-specific preferences and addressing cross-domain inconsistencies?}
    \item \textbf{Privacy-Utility Trade-offs:} \textit{How can federated learning frameworks balance privacy guarantees and model utility in CCTR prediction, minimizing noise impact while ensuring data confidentiality?}
\end{enumerate}

To address these open challenges, we present \textit{FedCCTR-LM}, a unified framework designed to harmonize augmentation, representation disentanglement, and adaptive privacy in federated, cross-domain environments. Our solution comprises three interlocking components. First, the \textit{Privacy-Preserving Augmentation Network} (PrivAugNet) harnesses large language models (LLMs) to enrich user and item features and to expand interaction sequences. This step substantially mitigates data sparsity, aligning domain-specific characteristics into a more consistent and informative feature space. Next, our \textit{Independent Domain-Specific Transformer with Contrastive Learning} (IDST-CL) module employs dedicated domain-specific transformers and contrastive objectives. Within IDST-CL, \textit{Intra-Domain Representation Alignment} (IDRA) refines augmented sequences by aligning them with original signals to reduce noise, while \textit{Cross-Domain Representation Disentanglement} (CDRD) segregates shared patterns from domain-unique traits. Collectively, these techniques ensure both stable intra-domain learning and effective cross-domain transfer. Finally, we introduce \textit{Adaptive Local Differential Privacy} (AdaLDP), a mechanism that dynamically adjusts noise levels to preserve privacy while safeguarding essential behavioral patterns.

The contributions of this work are summarized as follows:

\begin{itemize}
    \item[-] \textbf{A Paradigm Shift in Federated CCTR Prediction}: FedCCTR-LM is the first framework to seamlessly integrate CCTR prediction within a federated environment, fostering privacy-conscious collaboration across domains. 
    \item[-] \textbf{Innovative Augmentation Network}: We propose PrivAugNet, leveraging LLMs to counter data sparsity and heterogeneity, yielding richer and more consistent representations without centralizing raw data. 
    \item[-] \textbf{Advanced Representation Learning}: Our IDST-CL model disentangles domain-specific and cross-domain preferences through contrastive learning, enabling efficient knowledge transfer and addressing representational inconsistencies across domains. 
    \item[-] \textbf{Dynamic Privacy Mechanism}: Our AdaLDP mechanism dynamically balances privacy constraints and model utility, preserving sensitive user information while maintaining predictive fidelity. 
    \item[-] Extensive experiments on four benchmark datasets demonstrate the framework's superior CTR prediction accuracy, privacy compliance, and scalability, outperforming traditional, cross-domain, and federated learning baselines. 
\end{itemize}

The remainder of this paper is organized as follows. Section~\ref{sec:related_study} reviews related work on CCTR prediction, federated learning, and augmentation mechanisms. Section~\ref{sec:methodology} details the design and implementation of FedCCTR-LM, including its three core components. Section~\ref{sec:discussion} discusses implications and potential limitations, while Section~\ref{sec:experiments} presents experimental evaluations on real-world datasets. Finally, Section~\ref{sec:conclusion} concludes with insights and directions for future research.

\section{Related Study}
\label{sec:related_study}

\subsection{Cross-Domain CTR Prediction}
As users increasingly engage with a multitude of platforms—ranging from e-commerce marketplaces to streaming services—CCTR prediction has surfaced as a compelling challenge and a key opportunity to enhance personalization. Traditional CTR models, grounded in linear methods (e.g., logistic regression\citep{agarwal2015comparative}) or shallow feature-based frameworks (e.g., gradient-boosting decision trees\citep{di2022new}), initially provided workable solutions in single-domain contexts. Their limitations became apparent, however, as application scenarios expanded to diverse domains with heterogeneous item categories, feature distributions, and user behavior patterns. The introduction of deep learning models such as Wide\&Deep\citep{cheng2016wide}, DeepFM\citep{guo2017deepfm}, DIEN\citep{zhou2019deep}, and BST\citep{chen2019behavior} brought improved capacity for capturing non-linear feature interactions and modeling temporal dynamics within a single domain, but still fell short when faced with the broader complexities of multi-domain ecosystems.

The central premise of CCTR research is that knowledge gleaned from one (or multiple) well-populated, relatively stable “source” domains can inform and enrich CTR predictions in a sparser or emerging “target” domain\citep{zang2022survey}. Early attempts employed joint learning paradigms and transfer learning strategies. For instance, frameworks like STAR\citep{sheng2021one} and MiNet\citep{ouyang2020minet} sought to map user representations from a source domain to a target domain, leveraging auxiliary information—such as textual item descriptions or historical behavioral patterns—to alleviate data sparsity. Dual-transfer architectures like CoNet\citep{hu2018conet} introduced bidirectional parameter sharing, striving to balance shared semantic structures with domain-specific adaptations. Similarly, iterative transfer models like DDTCDR\citep{li2020ddtcdr} progressively aligned source and target domain representations, deepening cross-domain interactions over multiple training steps. Yet, these solutions often struggled with several persistent challenges. For example, conflicting optimization objectives across domains could lead to gradient interference, where improvements in one domain detracted from performance in another\citep{zang2022survey}. Additionally, many models relied on homogeneously defined feature spaces or assumptions about the “affluent” vs. “sparse” domain roles, oversimplifying real-world scenarios where domain boundaries and data richness levels are fluid and evolving over time\citep{zang2022survey}.

To more effectively reconcile domain heterogeneity, recent lines of inquiry have focused on sophisticated representation learning and alignment techniques. Contrastive learning\citep{cao2022c2dsr,ma2024triple,ye2023dream,zang2023contrastive}, in particular, has advanced the state-of-the-art by providing a principled approach to align user-item embeddings across domains without forcing rigid structural assumptions. Models like CMVCDR\citep{zang2023contrastive} integrate multi-view contrastive objectives to align static and sequential user preferences, bridging semantic gaps that arise from domain-specific terminologies or divergent item attribute distributions. DREAM\citep{ye2023dream} and related decoupled representation learning frameworks incorporate supervised contrastive tasks that disentangle shared and domain-specific features, forging embeddings that are both more robust and adaptable. Graph-based methods, such as C$^2$DSR\citep{cao2022c2dsr}, further extend this notion by integrating graph neural networks with sequential modeling and mutual information maximization, capturing high-order relational dependencies that transcend individual domain boundaries. Concurrently, temporal and sequence-centric solutions like TriCDR\citep{ma2024triple} incorporate cross-domain attention and multi-level contrastive learning to address temporal misalignments, domain overlaps, and the complex interplay between long-tail user behaviors and item availability.

Despite these advancements, existing CCTR methods face significant challenges, including reliance on centralized architectures that conflict with privacy constraints, and difficulty balancing domain-specific personalization with effective cross-domain knowledge transfer. Moreover, many existing approaches rely on assumptions of cross-domain feature homogeneity and pre-defined distinctions between affluent and sparse domains, which fail to account for the complexity and variability of real-world data distributions, resulting in negative transfer or biased representations.

\subsection{Augmentation Mechanisms}
Data augmentation has long been a pivotal strategy for enhancing model robustness and addressing challenges such as data sparsity, noise contamination, and feature incompleteness\citep{liu2023federated}. Traditional methods, including Singular Value Decomposition (SVD)\citep{parida2023svd} and Autoencoders (AEs)\citep{islam2021crash}, focus on decomposing or projecting data into more informative representations. SVD isolates core signal components by factorizing data into singular values and vectors, while AEs compress and reconstruct inputs in lower-dimensional spaces, extracting latent features and filtering extraneous noise. In federated scenarios, these classical techniques help ensure that local models discard irrelevant or low-quality inputs contributed by heterogeneous clients, thereby stabilizing training dynamics\citep{fontenla2021dsvd}. However, their utility in CCTR prediction remains limited: static, domain-agnostic transformations cannot easily accommodate the interplay of distinct domains, nor can they capture subtle semantic cues essential for transferring knowledge between them. Moreover, the complexity of adhering to strict privacy requirements in multi-domain, federated environments further constrains the effectiveness of these conventional methods.

The integration of LLMs into data augmentation pipelines has recently emerged as a powerful tool for enriching user-item interactions\citep{wei2024llmrec}, particularly in privacy-sensitive, cross-domain contexts. LLM-driven augmentation distinguishes itself by generating contextually meaningful, semantically rich expansions of sparse or incomplete datasets. Models such as RLMRec\citep{ren2024representation}, GraphGPT\citep{tang2024graphgpt}, and LLMRec\citep{wei2024llmrec} leverage pre-trained language models to synthesize additional textual attributes, item descriptors, or user preference signals that better align heterogeneous feature spaces across domains. This semantic enrichment not only bolsters intra-domain modeling by filling in gaps left by scarce or noisy inputs but also strengthens cross-domain transfer by elucidating patterns and relationships obscured by traditional augmentation techniques. Nevertheless, deploying LLM-based augmentation in federated CCTR systems presents challenges, including the computational overhead of fine-tuning or querying LLMs on-device and the risk of introducing irrelevant or misleading synthetic features if model prompts are not meticulously designed\citep{kuang2024federatedscope}. Balancing augmentation richness against noise and resource constraints requires careful calibration.

\subsection{Federated Learning}
FL \citep{mcmahan2017communication} has emerged as a transformative paradigm for privacy-preserving model training, allowing collaborative intelligence to be distilled from geographically or institutionally dispersed data sources without ever centralizing raw user interactions\citep{kairouz2021advances, zhang2021survey}. Early explorations, such as FCMF\citep{yang2021fcmf} and FedNCF\citep{perifanis2022federated}, demonstrated the feasibility of federated recommendation by applying secure aggregation protocols and carefully orchestrating local training steps, thereby preserving data confidentiality while uncovering essential user-item interaction patterns. Further refinements like FedCTR\citep{wu2022fedctr} introduced a federated native ad CTR prediction framework capable of leveraging cross-platform user behavior data and achieves effective user-interest modeling while preserving user privacy. These foundational efforts have since expanded into more complex, cross-domain recommendation scenarios. FedCDR\citep{yan2022fedcdr}, for instance, employs a client-server architecture to aggregate parameters across distinct domains, ensuring secure knowledge transfer without centralizing raw interactions. Similarly, Win-Win\citep{chen2023win} introduces a peer-to-peer framework that facilitates bidirectional domain collaboration, distributing benefits more equitably and employing differential privacy to preserve data confidentiality. FedDCSR\citep{zhang2024feddcsr}, in turn, pioneers disentangled representation learning within a federated context, enhancing both intra- and inter-domain feature representations to maintain robust CTR estimation across diverse data distributions. PPCDR\citep{tian2024privacy} integrates graph neural networks into the federated pipeline, adeptly handling structural complexity and fine-grained user preferences while minimizing communication overhead. 

Alongside architectural progress, robust privacy-preserving mechanisms have become integral to FL’s broader adoption\citep{kairouz2021advances, zhang2021survey}. While cryptographic techniques such as homomorphic encryption offer formal guarantees, their computational overhead often proves prohibitive for large-scale, latency-sensitive applications. Differential Privacy (DP) has thus gained favor as a more tractable solution, injecting calibrated noise into model updates to limit information leakage. Foundational work by McSherry et al.\citep{mcsherry2009differentially} and subsequent adaptations in federated matrix factorization contexts validate DP’s practicality in securing model training without wholesale performance collapse\citep{ribero2022federating}. Yet, current DP approaches frequently treat the privacy-utility trade-off as static, imposing uniform noise levels that can obliterate subtle but critical behavioral signals. This shortcoming is especially problematic in CTR prediction—where nuanced sequential patterns and domain-specific embeddings are central to predictive accuracy—and more recent methods like DS-ADMM++\citep{zhang2021ds}, despite balancing noise and performance within fixed frameworks, still fail to adaptively tune privacy measures as data distributions, user participation, and model sensitivity evolve. In essence, the future of federated learning for CCTR prediction hinges on devising adaptive privacy mechanisms that safeguard user data while preserving the delicate informational patterns that drive accurate recommendations. 

\section{Methodology}
\label{sec:methodology}

\begin{figure*}[!t]
    \centering
    \includegraphics[trim={0cm 20.66cm 0cm 0cm}, clip, width=\textwidth]{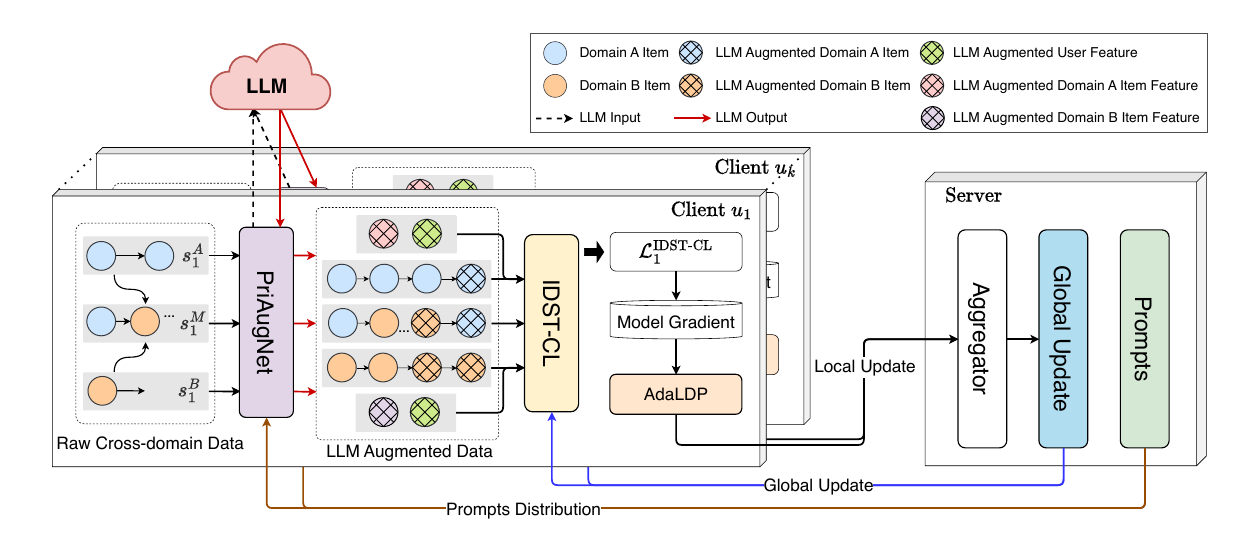}
    \caption{The architecture of the proposed FedCCTR-LM framework for federated cross-domain click-through rate (CTR) prediction. This framework integrates three key modules: the Privacy-preserving Augmentation Network (PrivAugNet), the Independent Domain-Specific Transformer with Contrastive Learning (IDST-CL), and the Adaptive Local Differential Privacy (AdaLDP) mechanism, collectively enabling effective privacy-preserving CCTR prediction.}
    \label{fig:fedcctr_lm_framework}
\end{figure*}

In this section, we formalize the CCTR prediction problem within a federated learning framework. Subsequently, we provide a comprehensive overview of the proposed \textit{FedCCTR-LM} methodology, elaborating on its three core components: \textit{Privacy-preserving Augmentation Network (PrivAugNet)}, \textit{Independent Domain-Specific Transformer with Contrastive Learning (IDST-CL)}, and \textit{Adaptive Local Differential Privacy (AdaLDP)}. Finally, the federated learning process for model optimization is discussed.

\begin{table}[width=.9\linewidth,cols=2,pos=!t]
\centering
\caption{Notations and Descriptions}
\label{tab:notations}
\renewcommand{\arraystretch}{1.2} 
\begin{tabular*}{\tblwidth}{@{}LL@{}}
\toprule
\textbf{Notation} & \textbf{Description} \\ 
\midrule
\(A, B, M\) & Represent domain A, domain B, and the cross-domain (mixed domain) \\
\(\mathcal{U}\) & The set of users common to both domains A and B \\
\(u_k\) & The \(k\)-th user within the overlapping user set \(\mathcal{U}\) \\
\(\mathcal{V}^A, \mathcal{V}^B\) & The set of items in domains A and B, respectively \\
\(\mathcal{D}_k, \mathcal{D}_k^+\) & The local training dataset for user \(u_k\) and the LLM-augmented dataset for user \(u_k\) \\ 
\midrule
\(s_k^A, s_k^B, s_k^M\) & User \(u_k\)'s interaction sequences in domains A, B, and cross-domain \\
\(\mathcal{P}_{\text{item}}, \mathcal{P}_{\text{user}}, \mathcal{P}_{\text{seq}}\) & Pre-defined augmentation prompts for items, users, and sequences \\
\(\mathcal{A}u_k\) & LLM-augmented user feature representation for user \(u_k\) \\
\(v_i^A, v_i^B, v_i^M\) & The \(i\)-th item interaction in domains A, B, and cross-domain \\
\(\tilde{v}_i^A, \tilde{v}_i^B, \tilde{v}_i^M\) & The \(i\)-th LLM-augmented item interaction in domains A, B, and cross-domain \\
\(\tilde{s}_k^A, \tilde{s}_k^B, \tilde{s}_k^M\) & LLM-augmented interaction sequences for user \(u_k\) in domains A, B, and cross-domain \\
\(\mathcal{E}s_k^A, \mathcal{E}s_k^B\) & LLM-expanded positive interaction sequences for user \(u_k\) in domains A and B \\
\(\mathcal{N}s_k^A, \mathcal{N}s_k^B\) & LLM-generated negative interaction sequences for user \(u_k\) in domains A and B \\
\(\mathcal{A}s_k^A, \mathcal{A}s_k^B, \mathcal{A}s_k^M\) & Final augmented interaction sequences for user \(u_k\) in domains A, B, and cross-domain \\ 
\midrule
\(\mathbf{e}^A_{\text{side}}, \mathbf{e}^B_{\text{side}}\) & Side information embeddings for domains A and B, respectively \\
\(\mathbf{e}_{\text{item},i}^A, \mathbf{e}_{\text{item},i}^B, \mathbf{e}_{\text{item},i}^M\) & The \(i\)-th item embedding in domains A, B, and cross-domain \\
\(\mathcal{T}^A(\cdot), \mathcal{T}^B(\cdot), \mathcal{T}^M(\cdot)\) & Domain-specific transformers for domains A, B, and cross-domain \\
\(\mathbf{h}^A, \mathbf{h}^B, \mathbf{h}^M\) & Feature representations of LLM-augmented sequences in domains A, B, and cross-domain \\
\(\mathbf{h}'^A, \mathbf{h}'^B\) & Feature representations of original sequences in domains A and B\\ 
\midrule
\(\mathcal{F}^A, \mathcal{F}^B\) & Multi-layer perceptrons (MLPs) for CTR prediction in domains A and B \\
\(\mathcal{L}^\text{IDRA}\) & Loss function for Intra-Domain Representation Alignment \\
\(\mathcal{L}^\text{CDRD}\) & Loss function for Cross-Domain Representation Disentanglement \\
\(\mathcal{L}^A, \mathcal{L}^B\) & Binary cross-entropy losses for CTR prediction in domains A and B \\
\(\mathcal{L}^\text{IDST-CL}\) & Overall loss function for the IDST-CL model \\ 
\midrule
\(|\mathcal{C}|\) & Number of candidates items sampled for interaction sequence expansion \\
\(\alpha\) & Margin that controls the alignment strength \\
\(\tau\) & Temperature coefficient \\
\(\rho\) & Sample ratio of clients for each round \\
\(T\) & Number of training iterations \\
\(\eta\) & Learning rate for model updates \\
\(\Theta^\text{G}\) & Global model parameters \\
\(\nabla_{k,t}\) & Gradients of client \(u_k\) at training round \(t\) \\
\(\tilde{\nabla}_{k,t}\) & Noisy gradients of client \(u_k\) at training round \(t\) \\
\(\theta\) & Gradient clipping threshold \\
\(\mathcal{R}\) & Decay factor for noise standard deviation \\
\(\sigma_t\) & Noise standard deviation at round \(t\) \\
\(\zeta\) & Order of Rényi Differential Privacy \\
\(\epsilon\) & Privacy budget for differential privacy \\
\bottomrule
\end{tabular*}
\end{table}

\subsection{Problem Formulation}

Consider a set of users \(\mathcal{U}\), where each user \(u_k \in \mathcal{U}\) interacts with items in two distinct domains, \(A\) and \(B\). Let the historical interaction sequences for user \(u_k\) be:
\begin{equation}
    s_k^A = \{v_1^A, v_2^A, \dots, v_i^A\} \quad\text{and}\quad s_k^B = \{v_1^B, v_2^B, \dots, v_j^B\},
\end{equation}
where \(v_i^A\) and \(v_j^B\) denote items that user \(u_k\) interacted with in domains \(A\) and \(B\), respectively. To capture cross-domain behavioral patterns, we additionally define a cross-domain sequence:
\begin{equation}
    s_k^M = \{v_1^M, v_2^M, \dots, v_{i+j}^M\},
\end{equation}
which is formed by chronologically merging \(s_k^A\) and \(s_k^B\). The sequence \(s_k^M\) provides a unified view of the user's historical interactions across both domains.

The central objective of CCTR prediction is to estimate the probability that a user \(u_k\) will click on a candidate item, leveraging their interaction history from both domains. Specifically, for a candidate next item \(v_{i+1}^A\) in domain \(A\), we aim to predict:
\begin{equation}
\hat{y}_k^A = \mathbb{P}(y_{i+1}^A = 1 \mid s_k^A, s_k^B, s_k^M),
\end{equation}
where \(y_{i+1}^A\) is a binary indicator of whether the user clicks on the next candidate item in domain \(A\). Similarly, for domain \(B\):
\begin{equation}
\hat{y}_k^B = \mathbb{P}(y_{j+1}^B = 1 \mid s_k^B, s_k^A, s_k^M).
\end{equation}

To preserve user privacy, training is conducted under FL paradigm. Each user \(u_k\) holds private datasets \(\mathcal{D}_k^A\) for domain \(A\) and \(\mathcal{D}_k^B\) for domain \(B\), which remain on the local device. Each instance \((s_k^d, y_k^d) \in \mathcal{D}_k^d\) corresponds to a historical sequence \(s_k^d\) in domain \(d \in \{A,B\}\) and the observed binary click label \(y_k^d\). Let \(\mathcal{F}(\cdot; \Theta)\) be the prediction function parameterized by \(\Theta\).

The local loss for user \(u_k\) in domain \(d\) is defined as:
\begin{equation}
\mathcal{L}_k^d(\Theta; \mathcal{D}_k^d) = \frac{1}{|\mathcal{D}_k^d|}\sum_{(s_k^d, y_k^d) \in \mathcal{D}_k^d} \ell(\mathcal{F}(s_k^d; \Theta), y_k^d),
\end{equation}
where \(\ell(\cdot)\) is a suitable loss function (e.g., binary cross-entropy).

The global objective aggregates the local losses over all users and both domains:
\begin{equation}
\min_{\Theta} \ \mathcal{L} = \sum_{u_k \in \mathcal{U}} \bigl[\mathcal{L}_k^A(\Theta; \mathcal{D}_k^A) + \mathcal{L}_k^B(\Theta; \mathcal{D}_k^B)\bigr].
\end{equation}

During federated training, users perform local updates on \(\Theta\) using their on-device data and periodically send model updates to a central server for aggregation. This process iterates until convergence, yielding a globally optimized parameter set \(\Theta\) that effectively leverages cross-domain interactions without exposing raw user data. As a result, the proposed approach aims to enhance CTR prediction accuracy and robustness in a privacy-preserving manner.  

\subsection{Framework Overview}

In this subsection, we introduce the proposed FedCCTR-LM framework, which is meticulously engineered to address the challenges of \textit{Non-I.I.D. Data distributions}, \textit{Cross-Domain Knowledge Transfer}, and \textit{Privacy-Utility Trade-offs} within FL based CCTR paradigm. The comprehensive architecture of FedCCTR-LM is depicted in Figure~\ref{fig:fedcctr_lm_framework}, with key notations and descriptions enumerated in Table~\ref{tab:notations}. The subsequent subsections delve into the intricacies of each constituent component.

We commence with PrivAugNet, an data augmentation mechanism that leverages the generative prowess of LLMs to enhance both user and item features. This component mitigates user-level imbalanced data sparsity and improves feature consistency across domains. Next, we introduce the IDST-CL, which consists of independent domain-specific transformers that capture personalized preferences within each domain, augmented by contrastive learning strategies. Specifically, we delve into Intra-Domain Representation Alignment (IDRA), which ensures robust domain-specific representations, and Cross-Domain Representation Disentanglement (CDRD), which facilitates the alignment of cross-domain representations with domain-specific ones to strengthen cross-domain learning. Finally, we elucidate AdaLDP, which governs the federated model training process, enabling precise control over privacy-accuracy trade-offs based on the sensitivity of user data, thus ensuring privacy guarantees are upheld without significantly compromising model performance.

\subsection{Privacy-Preserving Augmentation Network}

\begin{figure*}[!t]
    \centering
    \includegraphics[width=\textwidth, trim={0 21cm 0 0.6cm}, clip]{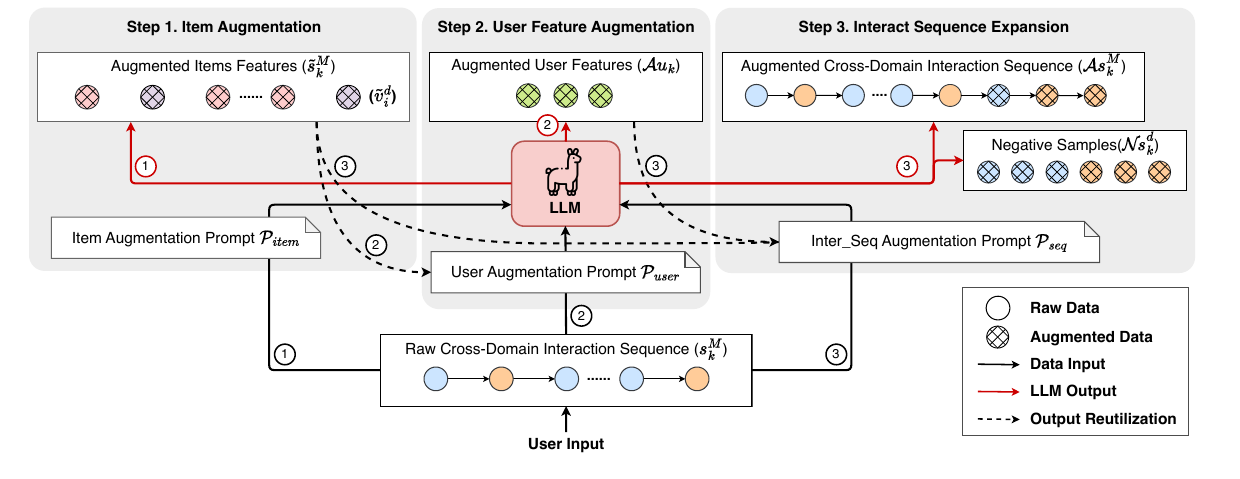}
    \caption{Overview of the Privacy-Preserving Augmentation Network (PrivAugNet) for enhancing CCTR prediction with LLM-based augmentation.}
    \label{fig:PrivAugNet}
\end{figure*}

PrivAugNet is designed to mitigate the challenges of user side data sparsity, feature incompleteness, and cross-domain inconsistency. Grounded in recent advances in prompt engineering\citep{kojima2022large, yang2024large,lee2024star, wei2024llmrec, ren2024representation} and LLMs\citep{chang2024survey}, PrivAugNet operates through three sequential stages: 1) Item Feature Augmentation, 2) User Profile Augmentation, and 3) Interaction Sequence Expansion. As shown in Figure~\ref{fig:PrivAugNet}, each stage utilizes specially crafted prompts to guide LLMs in generating augmented representations, creating richer and more consistent data for federated CCTR prediction. 

\subsubsection{Item Feature Augmentation}

\begin{figure}[!t]
    \centering
    \begin{subfigure}[b]{0.45\textwidth}
        \centering
        \includegraphics[width=\linewidth, trim={0 11.5cm 0 0}, clip]{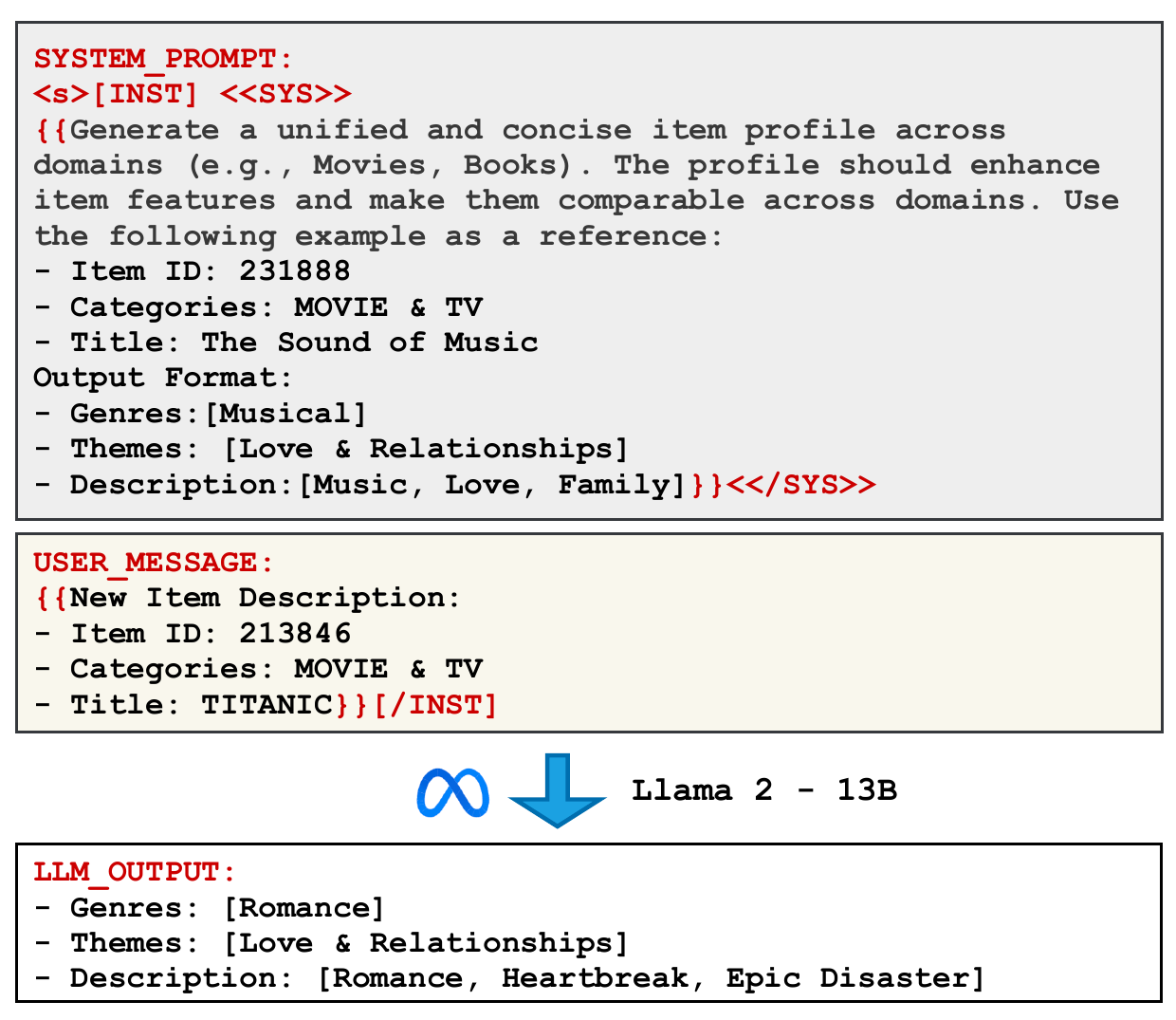}
        \caption{Item Feature Augmentation.}
        \label{fig:Item_Prompt}
    \end{subfigure}
    \hfill
    \begin{subfigure}[b]{0.45\textwidth}
        \centering
        \includegraphics[width=\linewidth, trim={0 11.5cm 0 0}, clip]{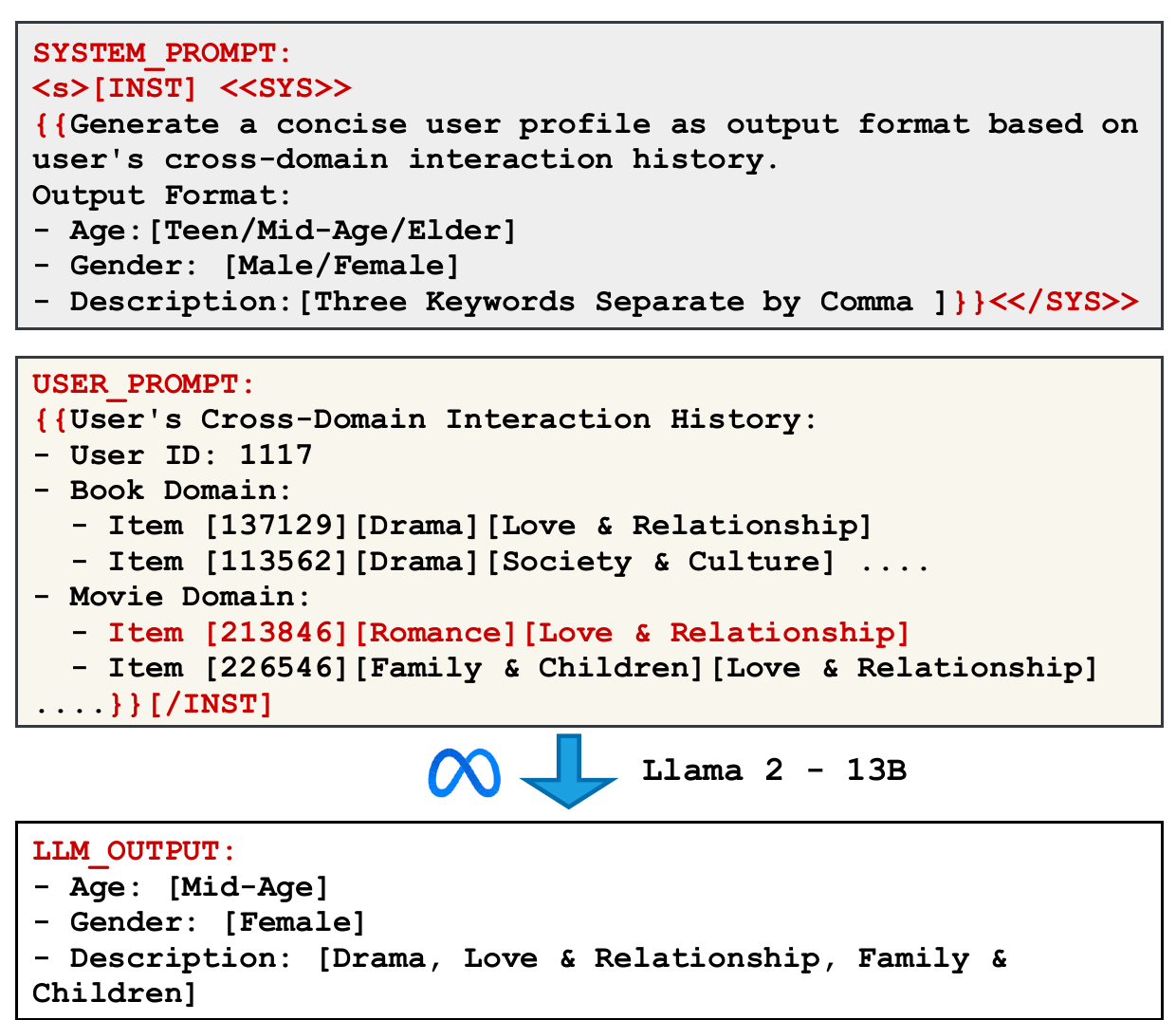}
        \caption{User Profile Augmentation.}
        \label{fig:User_Prompt}
    \end{subfigure}
    \vspace{1em} 
    \begin{subfigure}[b]{0.45\textwidth}
        \centering
        \includegraphics[width=\linewidth, trim={0 8.4cm 0 0}, clip]{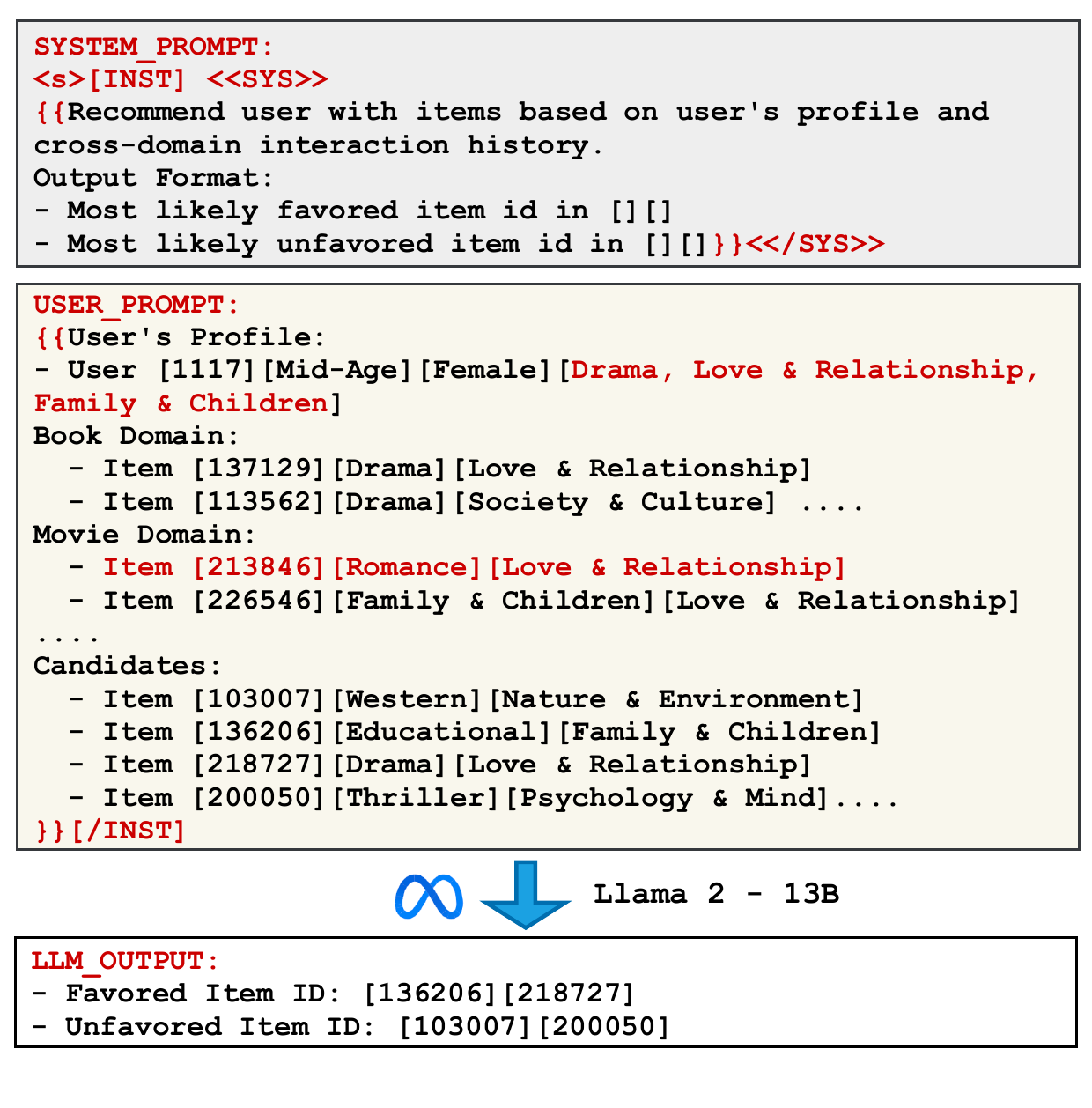}
        \caption{Interaction Sequence Expansion.}
        \label{fig:Seq_Prompt}
    \end{subfigure}
    \caption{Illustrations of the three steps of augmentation processes within PrivAugNet.}
    \label{fig:augmentation_processes}
\end{figure}

In the first stage, PrivAugNet focuses on augmenting item features to address feature incompleteness and improve cross-domain consistency. For a given user \(u_k\), whose local cross-domain interaction sequences are denoted as \(s_{k}^{M} = \{v_1^d, v_2^d, \ldots, v_i^d\},\ d\in\{A,B\}\), the augmentation process employs a predefined item augmentation prompt $\mathcal{P}_{\text{item}}$, as shown in Figure~\ref{fig:Item_Prompt}. This prompt is passed to the LLM to generate enriched item representations. The augmented item features are denoted as \(\tilde{v}_i^A\) and \(\tilde{v}_j^B\) for domains \(A\) and \(B\), respectively. The augmentation process can be formally expressed as:

\begin{equation}
\tilde{v}_i^d \leftarrow \text{LLM}(\mathcal{P}_\text{item}, v_i^d), \quad d\in\{A,B\}
\end{equation}

The Item Feature Augmentation harmonizes and enhances item features across both domains, making them more comparable. The augmented item representations $\tilde{v}_i^A$ and $\tilde{v}_j^B$ are then passed forward to the next stage for \textbf{User Profile Augmentation}, ensuring that the augmented item features contribute to enhancing the user features.

\subsubsection{User Profile Augmentation}

Following the augmentation of item features, PrivAugNet proceeds to augment the user profiles. This step aims to integrate cross-domain behavioral patterns by generating a comprehensive user representation \(\mathcal{A}u_k\) from the augmented interaction sequences \(\tilde{s}_k^M\). The LLM processes a user augmentation prompt \(\mathcal{P}_{\text{user}}\), generating an enriched user profile:

\begin{equation}
\mathcal{A}u_k \leftarrow \text{LLM}(\mathcal{P}_{\text{user}}, \tilde{s}_k^M)
\end{equation}

As illustrated in Figure~\ref{fig:User_Prompt}, the user profile is enriched by incorporating cross-domain behavior. The augmented user profile \(\mathcal{A}u_k\) encapsulates the user's preferences across both domains, improving CTR prediction accuracy. Importantly, the enhanced item features from the previous step contribute to the generation of a more accurate user profile, which subsequently feeds into the \textbf{Interaction Sequence Expansion} phase.

\subsubsection{Interaction Sequence Expansion}

The final stage of PrivAugNet builds upon the LLM-enhanced user and item features from the previous two steps. As shown in Figure~\ref{fig:Seq_Prompt}, the Interaction Sequence Expansion addresses data sparsity by expanding user interaction sequences and generating both positive and negative samples to augment the local training dataset. This process utilizes the interaction sequence augmentation prompt \(\mathcal{P}_{\text{seq}}\), directing the LLM to generate augmented interactions. Specifically, for each user \(k\), the LLM generates additional positive interactions \(\mathcal{E}s_k^A\) and \(\mathcal{E}s_k^B\), as well as negative samples \(\mathcal{N}s_k^A\) and \(\mathcal{N}s_k^B\) for domains \(A\) and \(B\), respectively:

\begin{equation}
    \mathcal{E}s_k^A, \mathcal{E}s_k^B, \mathcal{N}s_k^A, \mathcal{N}s_k^B \leftarrow \text{LLM}(\mathcal{P}_{\text{seq}}, \mathcal{A}u_k, \tilde{s}_k^M, \mathcal{C}^A, \mathcal{C}^B)
\end{equation}

Here, \(\mathcal{A}u_k\) and \(\tilde{s}_k^M\) are the user features and item features that have been enhanced by the PrivAugNet in the previous steps. The candidate item sets \(\mathcal{C}^A\) and \(\mathcal{C}^B\) are randomly sampled from the item pools \(\mathcal{V}^A\) and \(\mathcal{V}^B\) of domains \(A\) and \(B\), respectively, with the sample size determined by the hyperparameter \(|\mathcal{C}| = |\mathcal{C}^A| = |\mathcal{C}^B|\). These candidate items represent potential interactions for further augmentation.

The augmented positive interactions \(\mathcal{E}s_k^A\) and \(\mathcal{E}s_k^B\) are then integrated with the original interaction sequences to form enriched user interaction sessions:

\begin{equation}
\mathcal{A}s_k^{A} = \tilde{s}_k^{A} \cup \mathcal{E}s_k^A, \quad \mathcal{A}s_k^{B} = \tilde{s}_k^{B} \cup \mathcal{E}s_k^B
\end{equation}

These enriched sequences \(\mathcal{A}s_k^A\) and \(\mathcal{A}s_k^B\) offer a more comprehensive representation of the user's preferences, addressing data sparsity issues. The negative sequences \(\mathcal{N}s_k^A\) and \(\mathcal{N}s_k^B\) are used to form negative training sets, improving the model's ability to differentiate between relevant and irrelevant items.

\subsection{IDST-CL}

\begin{figure*}[!t]
\centering
    \includegraphics[width=\textwidth, trim=0cm 21.5cm 0cm 0.6cm, clip]{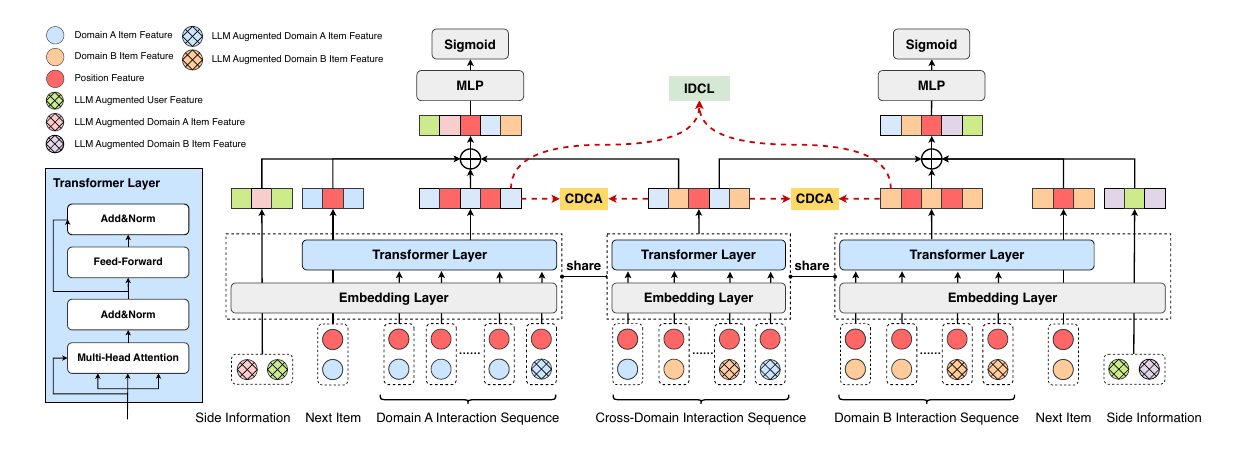}
    \caption{Overview of the IDST-CL Model: Independent Domain-Specific Transformer with Contrastive Learning. The model integrates independent domain-specific transformers for personalized representation learning, combined with contrastive learning techniques to enhance cross-domain knowledge transfer and alignment while preserving domain-specific personalization.}
    \label{fig:IDST-CL}
\end{figure*}

This subsection introduces the Independent Domain-Specific Transformer with Contrastive Learning (IDST-CL) model, which is specifically designed to address the challenges of CCTR prediction within the context of federated learning. The key challenge in this domain arises from the need to balance domain-specific personalization with effective knowledge transfer across domains. In particular, overlapping users who exhibit interaction sequences in both domains \(\mathcal{A}\) and \(\mathcal{B}\) require a model that can preserve unique domain behaviors while aligning shared preferences. To meet these requirements, the IDST-CL model employs independent transformers for each domain, along with contrastive learning techniques that facilitate cross-domain alignment while maintaining domain-specific personalization.

The design of the IDST-CL model is motivated by the Behavior Sequence Transformer (BST)\citep{chen2019behavior}, which has demonstrated effectiveness in modeling sequential dependencies in user interaction data. While BST focuses on modeling behaviors within a single domain, IDST-CL extends this concept by introducing independent transformers for each domain in the federated learning scenario. Specifically, the model incorporates domain-specific transformers—\(\mathcal{T}^A\) for domain A, \(\mathcal{T}^B\) for domain B, and a shared transformer \(\mathcal{T}^M\) for cross-domain interaction sequences. This architecture allows for the independent modeling of user behaviors in each domain while simultaneously learning shared representations that enhance cross-domain knowledge transfer.

\subsubsection{Embedding Layer}

The first component of the IDST-CL model is the embedding layer, which transforms various input features—such as user profiles, item features, and interaction sequences—into fixed-size, low-dimensional vectors. These vectors are subsequently used as input for the transformer layers. Side information, such as user and item metadata, is embedded into low-dimensional vectors using domain-specific embedding matrices. For instance, in domain \(\mathcal{A}\), side information is embedded as \(\mathbf{e}^A_\text{side}\) using the matrix \(\mathbf{W}_{\text{other}}^A \in \mathbb{R}^{|\mathcal{D}^A| \times d_o}\), where \(|\mathcal{D}^A|\) represents the size of the feature space and \(d_o\) is the embedding dimension. A similar embedding process is applied for domain \(\mathcal{B}\).

For each item in the interaction sequence, both item features (such as item ID and LLM-augmented attributes) and positional features are embedded. For domain \(\mathcal{A}\), the item embeddings are constructed using \(\mathbf{W}_{\text{item}}^A \in \mathbb{R}^{|\mathcal{V}^A| \times d_v}\), where \(d_v\) is the embedding dimension, and positional encoding are applied separately. The final item representation is obtained by concatenating the item ID, item features, and positional encoding:

\begin{equation}
    \mathbf{e}_{\text{item},i}^A = [\mathbf{e}_{\text{id},i}^A \oplus \mathbf{e}_{\text{feat},i}^A \oplus \mathbf{e}_{\text{pos},i}^A] \in \mathbf{R}^{d_v}
\end{equation}

For cross-domain sequences, the same embedding process is applied to ensure consistency across domains:

\begin{equation}
    \mathbf{e}_{\text{item},i}^M = [\mathbf{e}_{\text{id},i}^M \oplus \mathbf{e}_{\text{feat},i}^M \oplus \mathbf{e}_{\text{pos},i}^M] \in \mathbf{R}^{d_v}
\end{equation}

\subsubsection{Transformer Layer}

We utilize self-attention mechanism to capture dependencies both within and between items in each domain's sequence. For domain \(\mathcal{A}\), the input is the sequence embedding \(\mathbf{E}^A = \{\mathbf{e}_{\text{item},1}^A, \mathbf{e}_{\text{item},2}^A, \dots, \mathbf{e}_{\text{item},i}^A\}\), which is processed through the query, key, and value matrices in the scaled dot-product attention mechanism:

\begin{equation}
    \text{Attention}(Q, K, V) = \text{softmax}\left(\frac{Q K^\top}{\sqrt{d_k}}\right)V
\end{equation}

where \(Q\), \(K\), and \(V\) are the query, key, and value matrices derived from the input embeddings. This process is repeated for domain \(\mathcal{B}\) and the cross-domain sequence \(\mathcal{T}^B\) and \(\mathcal{T}^M\), respectively.

The multi-head attention mechanism captures diverse interaction patterns within each domain and across domains, and the results are concatenated and linearly transformed:

\begin{equation}
    \mathbf{S}^A = \text{Concat}(\text{head}_1,\, \dots,\, \text{head}_H)\mathbf{W}_O
\end{equation}

Following the multi-head attention, residual connections and layer normalization are applied, followed by a position-wise Feed-Forward Network (FFN):

\begin{equation}
    \mathbf{F}^A = \text{ReLU}(\mathbf{S}'^A \mathbf{W}_1 + \mathbf{b}_1)\mathbf{W}_2 + \mathbf{b}_2
\end{equation}

This process is also applied to domain \(\mathcal{B}\) and cross-domain interaction sequences to yield \(\mathbf{F}^B\) and \(\mathbf{F}^M\), respectively. Residual connections and normalization are again applied to the outputs, and mean pooling is performed across the sequence length to obtain fixed-size vectors:

\begin{equation}
        \mathbf{h}^A = \text{Pooling}(\mathbf{F}'^A),\quad \mathbf{h}^B = \text{Pooling}(\mathbf{F}'^B),\quad \mathbf{h}^M = \text{Pooling}(\mathbf{F}'^M)
\end{equation}

\subsubsection{Intra-Domain Representation Alignment}

\begin{figure}[!t]
    \centering
    \begin{subfigure}[b]{0.48\textwidth}
        \centering
         \includegraphics[width=\linewidth, trim={0 23cm 5.5cm 0.5cm}, clip]{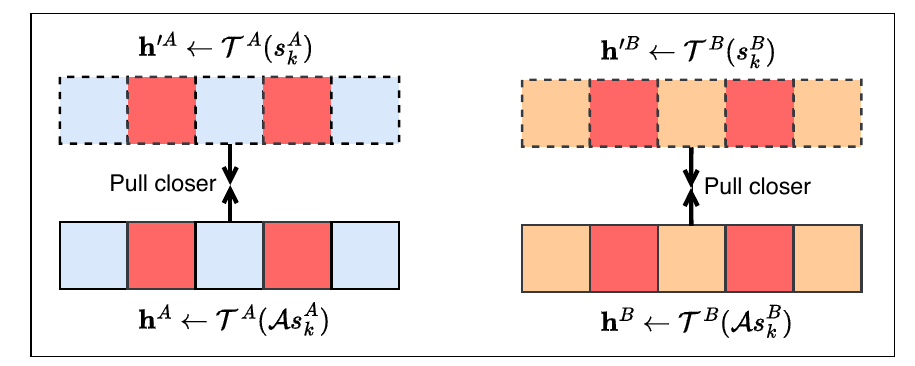}
         \caption{Illustration of the IDRA module, showing the alignment of augmented and original sequences within each domain to reduce noise introduced by LLM-augmented data}
         \label{fig:IDRA}
    \end{subfigure}
    \hfill
    \begin{subfigure}[b]{0.48\textwidth}
        \centering
        \includegraphics[width=\linewidth, trim={0 23cm 5.5cm 0.5cm}, clip]{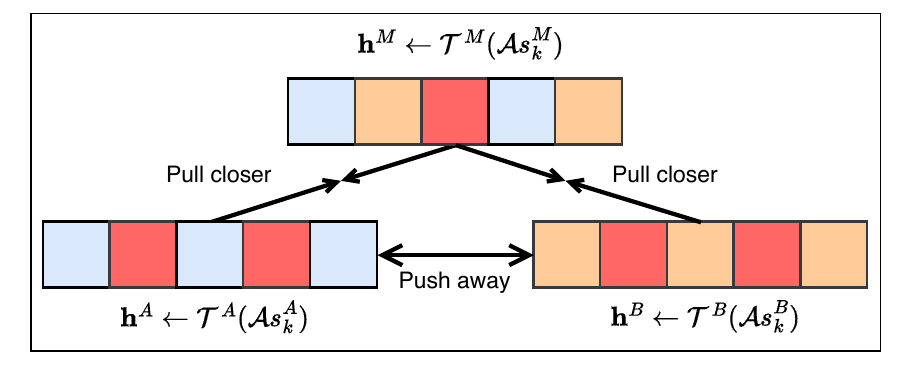}
        \caption{Illustration of the CDRD module, demonstrating the alignment of cross-domain representations with domain-specific ones while maintaining their independence.}
        \label{fig:CDRD}
    \end{subfigure}
    \caption{Visual representations of the contrastive learning within IDST-CL}
    \label{fig:contrastive learning}
\end{figure}

The integration of data generated through LLMs by PrivAugNet is crucial in alleviating data sparsity issues in user interaction sequences. However, this augmentation process inherently introduces noise\citep{wei2024llmrec}, as the synthetic interactions may not fully align with the authentic user preferences. The misalignment between the original and augmented sequences can distort domain-specific representations, potentially undermining the model’s ability to accurately capture user behaviors across domains.

To tackle this issue, we introduce the Intra-Domain Representation Alignment (IDRA) module, depicted in Figure~\ref{fig:IDRA}. The core objective of IDRA is to ensure that the augmented and original interaction sequences remain closely aligned within each domain. Specifically, for domain \(\mathcal{A}\), the augmented sequence \(\mathcal{A}s^A_k\) and the original sequence \(s^A_k\) should have corresponding representations \(\mathbf{h}^A_i\) and \(\mathbf{h}'^A_i\), respectively, while for domain \(\mathcal{B}\), the same alignment is sought between \(\mathcal{A}s^B_k\) and \(s^B_k\), with their respective representations \(\mathbf{h}^B_i\) and \(\mathbf{h}'^B_i\). By minimizing the misalignment between these augmented and original representations, IDRA reduces the noise introduced by the augmentation process and enhances the model’s robustness and accuracy.

The loss function for IDRA, applied to user \(u_k\), is defined as:

\begin{equation}
\label{eq:loss_IDRA}
        \mathcal{L}^\text{IDRA}_k = \frac{1}{N}\sum_{i\in\mathcal{D}_k}( \max(0,\alpha-\text{sim}(\mathbf{h}^A_i,\mathbf{h}'^A_i)) + \max(0,\alpha-\text{sim}(\mathbf{h}^B_i,\mathbf{h}'^B_i)))
\end{equation}

where \(\mathcal{D}_k\) represents the local training dataset for user \(u_k\), and \(\text{sim}(\mathbf{a}, \mathbf{b})\) denotes the cosine similarity between vectors \(\mathbf{a}\) and \(\mathbf{b}\). The hyperparameter \(\alpha\) defines the margin that controls the strength of the alignment. The loss function (Eq.\ref{eq:loss_IDRA}) aims to align the augmented and original representations within each domain, ensuring that they capture complementary aspects of the user’s behavior. While the augmented sequence may introduce noise, the alignment encourages the representations to stay within a reasonable range of each other, allowing both sequences to contribute to a richer, more robust domain-specific representation. The margin (controlled by \(\alpha\)) helps ensure that the augmented and original representations are sufficiently close to each other to prevent drastic divergence, yet still allows the augmented sequence to bring in additional information. This balance ensures that the model benefits from the enhanced sequences without losing the diversity introduced by the augmentation process.

\subsubsection{Cross-Domain Representation Disentanglement}

While the IDRA addresses the alignment of augmented and original sequences within each domain, it is equally crucial to disentangle the shared representations between domains \(\mathcal{A}\) and \(\mathcal{B}\) without compromising the unique characteristics of each domain. The challenge here is to capture the commonalities between user behaviors across domains while ensuring that domain-specific preferences are preserved\citep{zhang2024feddcsr}.

To tackle this challenge, we introduce the Cross-Domain Representation Disentanglement (CDRD) module, which is designed to decouple shared cross-domain representations from domain-specific features. The CDRD module ensures that the cross-domain representation \(\mathbf{h}_i^M\) is aligned with both domain \(\mathcal{A}\) and domain \(\mathcal{B}\) representations (\(\mathbf{h}_i^A\) and \(\mathbf{h}_i^B\)) to capture shared information, while simultaneously maintaining a distinct separation between \(\mathbf{h}_i^A\) and \(\mathbf{h}_i^B\). This disentanglement approach ensures that common behavioral patterns across domains are captured effectively, while also preserving the unique preferences of each domain, thus maintaining the personalizations required for robust CTR prediction.

As shown in Figure~\ref{fig:CDRD}, the cross-domain representation \(\mathbf{h}_i^M\) serves as a bridge to align the shared user behavior, but it is constrained to remain independent of domain-specific representations. The disentanglement process increases the distance between the representations of domain \(\mathcal{A}\) and domain \(\mathcal{B}\) (i.e., \(\mathbf{h}_i^A\) and \(\mathbf{h}_i^B\)), which reinforces the individuality of each domain’s behavioral patterns while allowing for cross-domain knowledge transfer.

The loss function for CDRD, as applied to user \(u_k\), is defined as follows:

\begin{equation}
        \mathcal{L}^{\text{CDRD}}_k = -\frac{1}{N} \sum_{i \in \mathcal{D}_k} \log \left( \frac{\exp{\left(\text{sim}(\mathbf{h}_i^M, \mathbf{h}_i^A)/\tau\right)} + \exp{\left(\text{sim}(\mathbf{h}_i^M, \mathbf{h}_i^B)/\tau\right)}}{\sum_{j \in \mathcal{D}_k} \exp{\left(\text{sim}(\mathbf{h}_j^A, \mathbf{h}_j^B)/\tau\right)}}\right)
\end{equation}

where \(\text{sim}(\mathbf{a}, \mathbf{b})\) denotes the cosine similarity between vectors \(\mathbf{a}\) and \(\mathbf{b}\), \(\mathcal{D}_k\) represents the local training dataset for user \(u_k\), and \(\tau\) is the temperature coefficient. This loss function encourages cross-domain representations to align with both domain-specific representations, while the denominator ensures that domain-specific information remains distinct by contrasting cross-domain similarities.

\subsubsection{Optimization Objectives}

For each domain, the final representations \(\mathbf{h}^A\) and \(\mathbf{h}^B\) are obtained by applying mean pooling to the output of the domain-specific transformers. These domain-specific representations are then concatenated with the shared cross-domain representation \(\mathbf{h}^M\) and the relevant side information \(\mathbf{e}_{\text{side}}\). The concatenated features are passed through domain-specific multi-layer perceptrons (MLPs) to predict the CTR for each domain. The process of concatenation and prediction through the MLPs can be expressed as:

\begin{equation}
        \hat{y}^A = \mathcal{F}^A([\mathbf{h}^A\oplus \mathbf{h}^M\oplus \mathbf{e}_{\text{side}}^A]), \quad\hat{y}^B= \mathcal{F}^B([\mathbf{h}^B\oplus \mathbf{h}^M\oplus \mathbf{e}_{\text{side}}^B])
\end{equation}

where \(\oplus\) denotes the concatenation operation. \(\mathcal{F}^A\) and \(\mathcal{F}^B\) are MLPs, each comprising three fully connected layers with ReLU activation functions, followed by a final sigmoid layer to output the predicted CTR probabilities for each domain.

For each user \(u_k\), the binary cross-entropy loss is calculated separately for domains \(\mathcal{A}\) and \(\mathcal{B}\) to evaluate the prediction performance in both domains. The loss functions for domain \(\mathcal{A}\) and domain \(\mathcal{B}\) are defined as:

\begin{equation}
    \mathcal{L}^A_k = -\frac{1}{N_k} \sum_{(x^A,y^A)\in \mathcal{D}_k}[y^A \log(\hat{y}^A)+(1-y^A)\log(1-\hat{y}^A)]
\end{equation}

\begin{equation}
    \mathcal{L}^B_k = -\frac{1}{N_k} \sum_{(x^B,y^B)\in \mathcal{D}_k}[y^B \log(\hat{y}^B)+(1-y^B)\log(1-\hat{y}^B)]
\end{equation}

where \(\mathcal{D}_k\) represents the local cross-domain training dataset for user \(u_k\). The variables \(x^A\) and \(x^B\) represent the input features, which include the target items \(v^A_{i+1}\) and \(v^B_{j+1}\) as well as the user's behavioral interaction sequences. The labels \(y^A\) and \(y^B\) are binary values indicating the click behavior for domains \(\mathcal{A}\) and \(\mathcal{B}\), respectively, where \(y^A, y^B \in \{0, 1\}\).

The overall optimization objective for user \(u_k\) in the federated learning setting combines the CTR prediction loss with the contrastive learning loss from the IDST-CL framework. The total loss is given by:

\begin{equation}
    \mathcal{L}_k^{\text{IDST-CL}}=\mathcal{L}_k^\text{IDST} + \mathcal{L}_k^\text{CL} = \mathcal{L}_k^A + \mathcal{L}_k^B + \lambda_1\mathcal{L}^\text{IDRA}_k + \lambda_2\mathcal{L}^\text{CDRD}_k
\end{equation}

where \(\lambda_1\) and \(\lambda_2\) serve to regulate the contribution of the IDRA and CDRD terms to the overall loss function. By tuning these parameters, the model can optimize the trade-off between enhancing domain-specific representations and ensuring effective cross-domain alignment, thereby refining the overall performance.

\subsection{Federated Model Training}

This subsection outlines the training process of the FedCCTR framework, which facilitates the PrivAugNet and IDST-CL model in a federated learning paradigm while preserving user privacy through AdaLDP. The training procedure operates in a decentralized fashion, with the server coordinating global model updates and client devices performing local computations on their respective data. The AdaLDP mechanism is introduced to dynamically adjust the noise added to gradients, ensuring privacy-preserving model updates while minimizing performance degradation.

\subsubsection{Overall Training Process}

\begin{algorithm}[htbp]
\SetAlgoNlRelativeSize{-1} 
\SetNlSty{}{}{:} 
\SetAlgoNlRelativeSize{0.5} 

\caption{FedCCTR-LM Training Process}
\KwSty{\textbf{SERVER-SIDE EXECUTION (\textit{FedServer})}:} \\
\KwIn{Number of iterations \(T\), learning rate \(\eta\), client list \(\mathcal{U}\)} 
\KwOut{Global model parameters \(\Theta_t^\text{G}\)} 
1. Initialize global model parameters \(\Theta_0^\text{G}\)\;
2. Broadcast LLM prompts \(\mathcal{P}_\text{item}\), \(\mathcal{P}_\text{user}\), \(\mathcal{P}_\text{seq}\) to all clients in \(\mathcal{U}\)\;
3. \For{each round \(t = 0, 1, \dots, T-1\)}{
    Sample a subset of clients \(\mathcal{U}_t \subset \mathcal{U}\)\;
    Broadcast global model parameters \(\Theta_t^\text{G}\) to all clients in \(\mathcal{U}_t\)\;
    \For{each client \(u_k \in \mathcal{U}_t\)}{
        Receive noisy gradient \(\tilde{\nabla}_{k,t}\) from client \(u_k\)\;
    }
    Update global model parameters using aggregated gradients as in Eq.(\ref{eq:global_update})
}
4. Broadcast final global model parameters \(\Theta_T^\text{G}\) to all clients\;

\vspace{0.5em}

\KwSty{\textbf{CLIENT-SIDE EXECUTION (\textit{User Device \(u_k\)})}:} \\
\KwIn{Privacy budget \(\epsilon\), decay rate \(\mathcal{R}\), noise standard deviation \(\sigma\), sample ratio \(\rho\), local dataset \(\mathcal{D}_k\)} 
\KwOut{Noisy gradient \(\tilde{\nabla}_{k,t}\)} 
1. Receive LLM prompts \(\mathcal{P}_\text{item}\), \(\mathcal{P}_\text{user}\), \(\mathcal{P}_\text{seq}\)\;
2. Generate augmented local training dataset \(\mathcal{D}_k^+\)\;
3. Initialize privacy budget \(\epsilon_0\) and noise standard deviation \(\sigma_0\)\;
4. \For{each round \(t = 0, 1, \dots, T-1\)}{
    Receive global model parameters \(\Theta_t^\text{G}\) from the server\;
    Compute local model loss:\(\mathcal{L}_{k,t}^\text{IDST-CL}(\Theta_t^\text{G}; \mathcal{D}_k^+)\)\;
    Compute local gradient:\(\nabla_{k,t} \gets \frac{\partial \mathcal{L}_{k,t}}{\partial \Theta_t^\text{G}}\)\;
    Apply AdaLDP to add noise:\(\tilde{\nabla}_{k,t} \gets \text{AdaLDP}(\nabla_{k,t}, \sigma_t, \epsilon_t, \rho)\)\;
    \If{received \textbf{Stop Participation} signal}{
        \textbf{break}\;
    }
    Send noisy gradient \(\tilde{\nabla}_{k,t}\) to the server\;
    Update noise standard deviation:\(\sigma_{t+1} \gets \sigma_t \cdot \mathcal{R}\)\;
}
\end{algorithm}

The training process for the FedCCTR-LM framework (outlined in \textbf{Algorithm 1}) begins with the \textbf{FedServer} initializing the global model parameters \(\Theta_0^\text{G}\) and broadcasting these parameters to all participating clients in set \(\mathcal{U}\), along with predefined prompts: the item prompt \(\mathcal{P}_{\text{item}}\), user prompt \(\mathcal{P}_{\text{user}}\), and sequence prompt \(\mathcal{P}_{\text{seq}}\).

On the client side, each user \(u_k\) utilizes these prompts to generate LLM-augmented item features \(\tilde{v}_i^A\), \(\tilde{v}_i^B\), and \(\tilde{v}_i^M\) for domains \(A\), \(B\), and cross-domain interactions, respectively. The LLM-augmented interaction sequences \(\mathcal{A}s_k^A\), \(\mathcal{A}s_k^B\), and \(\mathcal{A}s_k^M\) are then incorporated into the local dataset \(\mathcal{D}_k^+\). These expanded datasets support the local training of the IDST-CL, which entails computing the local loss \(\mathcal{L}_{k,t}(\Theta_t^\text{S};\mathcal{D}_k^+)\), where \(\mathcal{L}_{k,t}\) combines the CTR prediction losses \(\mathcal{L}^A\) and \(\mathcal{L}^B\) with the contrastive learning losses \(\mathcal{L}^\text{IDRA}\) and \(\mathcal{L}^\text{CDRD}\), as formalized in Sections 3.4.3 and 3.4.4.

Once the local loss is computed, each client derives the local gradients \(\nabla_{k,t}\) through back-propagation. To preserve user privacy, clients apply the AdaLDP mechanism to perturb the raw gradients with noise. This ensures that the final noisy gradients \(\tilde{\nabla}_{k,t}\), uploaded to the server, adhere to the client’s privacy budget \(\epsilon\). The server aggregates the noisy gradients from all clients and updates the global model parameters based on the FedAvg as (Eq.~\ref{eq:global_update}):

\begin{equation}
\Theta_t^\text{G} \leftarrow \Theta_{t-1}^\text{G} - \eta \frac{1}{|\mathcal{U}_t|} \sum_{u_k \in \mathcal{U}_t} \tilde{\nabla}_{k,t}
\label{eq:global_update}
\end{equation}

Here, \(\eta\) denotes the learning rate, and \(\mathcal{U}_t\) represents the subset of clients participating in round \(t\). The updated model \(\Theta_t^\text{G}\) is then broadcast back to all clients for the next round of training. This process repeats for \(T\) communication rounds, during which the global model progressively improves while ensuring user data remains private. The training stops after the global model converges or after a predefined number of rounds \(T\).

\subsubsection{Adaptive Local Differential Privacy}

\begin{algorithm}[htbp]
\caption{The AdaLDP Function \((\nabla_{k,t}, \sigma_t, \epsilon_t, \rho)\)}
\KwIn{Local gradient \(\nabla_{k,t}\), noise standard deviation \(\sigma_t\), privacy budget \(\epsilon_t\)}
\KwOut{Noisy gradient \(\tilde{\nabla}_{k,t}\) or \textbf{Stop Participation} signal}
1. Clip gradient \(\nabla_{k,t}\) using threshold \(\theta\) as Eq.(\ref{eq:gradient_clipping}) \\
2. Add Gaussian noise with standard deviation \(\sigma_t\) as Eq.(\ref{eq:noise_adding}) \\
3. Update privacy budget \(\epsilon_{t+1}\) by Eq.(\ref{eq:privacy_leakage}) and (\ref{eq:budget_update}) \\
4. \If{\(\epsilon_{t+1} \leq 0\)}{
    \Return \textbf{Stop Participation} signal\;
}
\Else{
    \Return Noisy gradient \(\tilde{\nabla}_{k,t}\)\;
}
\end{algorithm}

AdaLDP adaptively introduces noise to local gradients, beginning with high noise levels to protect sensitive information in early training rounds and gradually decreasing the noise via a decay factor \(\mathcal{R}\) as the model converges. This approach aligns the level of privacy protection with the gradient’s evolving sensitivity, enabling more informative model updates in later rounds.

The process starts by clipping the local gradient \(\nabla_{k,t}\) to limit sensitivity:

\begin{equation}
\nabla_{k,t} \leftarrow \nabla_{k,t} / \max\left(1, \frac{\|\nabla_{k,t}\|_2}{\theta}\right)
\label{eq:gradient_clipping}
\end{equation}

This ensures that no individual user’s data disproportionately affects the model. After clipping, Gaussian noise \(\mathcal{G}(0, \sigma_t^2 \mathbf{I})\) is added:

\begin{equation}
\tilde{\nabla}_{k,t} = \nabla_{k,t} + \mathcal{G}(0, \sigma_t^2 \mathbf{I})
\label{eq:noise_adding}
\end{equation}

The noise standard deviation \(\sigma_t\) is updated over time with the decay factor \(\mathcal{R}\):

\begin{equation}
\sigma_{t+1} = \sigma_t \cdot \mathcal{R}
\end{equation}

AdaLDP leverages Rényi Differential Privacy (RDP)\citep{mironov2017renyi} to monitor privacy leakage by accumulating the privacy budget \(\epsilon_t\) at each iteration:

\begin{equation}
\epsilon_t(\zeta) = \frac{1}{\zeta - 1} \ln\left(1 + \rho^2 \left( e^{(\zeta - 1)\frac{\theta^2}{\sigma_t^2}} - 1 \right) \right)
\label{eq:privacy_leakage}
\end{equation}

where \(\epsilon_t(\zeta)\) denotes the per-round privacy cost, \(\rho\) is the client sampling ratio, \(\theta\) is the clipping threshold, and \(\zeta\) is the order of RDP. The accumulated privacy expenditure is then updated:

\begin{equation}
\epsilon_{t+1} = \epsilon_t - \epsilon_t(\zeta)
\label{eq:budget_update}
\end{equation}

If the accumulated privacy budget \(\epsilon_{t+1}\) is less than or equal to zero, the client halts further participation in the training process to ensure privacy guarantees are maintained.

The proof of AdaLDP’s privacy guarantees is provided in \textbf{Appendix~\ref{sec:appendix_A}}, detailing how \((\epsilon, \delta)\)-differential privacy is ensured through adaptive noise injection and privacy budgeting. This mechanism provides a robust solution to maintaining user privacy while supporting effective model learning, thereby advancing the field of privacy-preserving federated learning.

\section{Discussion}
\label{sec:discussion}

\begin{table*}[!t]
\centering
\caption{Comparison of FedCCTR-LM with Existing CCTR Methods on Key Aspects}
\label{tab:discussion}
\resizebox{\textwidth}{!}{%
\begin{tabular}{lcccccc}
\toprule
\multirow{2}{*}{\textbf{Models}} & \multicolumn{2}{c}{\textbf{Data Utilization}} & \multirow{2}{*}{\textbf{Bidirectional Transfer}} & \multirow{2}{*}{\textbf{Privacy Preservation}} & \multirow{2}{*}{\textbf{CL Implementation}} \\
\cmidrule(lr){2-3}
 & \textbf{Mixed Sequence} & \textbf{LLM Augmentation} & & & \\
\midrule
CoNet\citep{hu2018conet} & \ding{55} & \ding{55} & \ding{55} & \ding{55} & \ding{55} \\
MiNet\citep{ouyang2020minet} & \ding{55} & \ding{55} & \ding{55} & \ding{55} & \ding{55} \\
RecGURU\citep{li2022recguru} & \ding{55} & \ding{55} & \ding{55} & \ding{55} & \ding{55} \\
FedCTR\citep{wu2022fedctr} & \ding{55} & \ding{55} & \ding{55} & Static DP & \ding{55} \\
C$^{2}$DSR\citep{cao2022c2dsr} & \ding{51} & \ding{55} & \ding{55} & \ding{55} & inter-CL \\
Tri-CDR\citep{ma2024triple} & \ding{51} & \ding{55} & \ding{55} & \ding{55} & Both intra- and inter-CL \\
FedDCSR\citep{zhang2024feddcsr} & \ding{51} & \ding{55} & \ding{55} & Static DP & \ding{55} \\
\midrule
\textbf{FedCCTR-LM (Ours)} & \ding{51} & \ding{51} & \ding{51} & Adaptive DP & Both intra- and inter-CL \\
\bottomrule
\end{tabular}%
}
\begin{tablenotes}
    \footnotesize
    \item \textbf{Note:}  \ding{51} indicates the feature is supported, \ding{55} indicates the feature is not supported.
\end{tablenotes}
\end{table*}

In this section, we present a comprehensive comparison between the proposed FedCCTR-LM framework and existing CCTR methods, focusing on key aspects such as domain data utilization, privacy preservation, contrastive learning (CL) implementation, and computational complexity. This comparison aims to highlight the unique contributions and enhanced effectiveness of FedCCTR-LM in addressing the challenges inherent in CCTR tasks. The key aspects comparison results are illustrated in Table~\ref{tab:discussion}. 

\subsection{Comparison with Existing CCTR Methods in Data Utilization}

CCTR prediction has traditionally centered on transferring user behavior data from a source domain to a target domain to improve CTR predictions. Early methods, such as CoNet \citep{hu2018conet} and MiNet \citep{ouyang2020minet}, leverage short-term or long-term user interests through selective cross-domain mappings, but they limit interactions to unidirectional flows. Later models like C$^2$DSR\citep{cao2022c2dsr} introduce contrastive learning to enhance cross-domain alignment, yet often rely on strict user alignment, which restricts adaptability in mixed-domain scenarios. Models such as RecGURU\citep{li2022recguru} generate global user representations to address sparsity but do not explicitly model mixed-domain sequences. 

In contrast, the proposed FedCCTR-LM introduces bidirectional domain utilization, incorporating domain $A$, domain $B$, and mixed domains without a fixed source-target designation. This flexibility is driven by PrivAugNet, which uses LLM-based augmentation to enrich sparse domains and ensure coherent cross-domain representations. Additionally, IDST-CL includes three independent transformers for domain $A$, domain $B$, and the mixed domain, autonomously capturing user interactions specific to each domain while supporting robust knowledge transfer. By integrating PrivAugNet and IDST-CL, FedCCTR-LM facilitates adaptive, privacy-preserving CCTR prediction, capturing user behavior nuances across domains with enhanced flexibility and predictive precision.

\subsection{Comparison with Existing CCTR Methods in Privacy Preservation}

Privacy preservation is a critical concern in CCTR prediction due to the sensitivity of user data shared across multiple digital environments. Traditional models like CoNet\citep{hu2018conet} and MiNet\citep{ouyang2020minet} expose user data to privacy risks through direct transfer of behavior patterns between domains. In contrast, FedCTR\citep{wu2022fedctr} and FedDCSR\citep{zhang2024feddcsr} adopt federated learning to avoid centralized data sharing, thus reducing potential exposure. However, their reliance on static noise injection throughout training presents challenges in adaptability and effectiveness. FedCTR’s static differential privacy mechanism can diminish model accuracy as training progresses.

In contrast, FedCCTR-LM leverages an AdaLDP mechanism to dynamically adjust noise based on data sensitivity and model convergence. AdaLDP provides stronger privacy protection during the early, most vulnerable training stages by applying higher noise levels and gradually reducing noise as the model stabilizes, balancing privacy with accuracy. This adaptive approach positions privacy as a fluid parameter that evolves alongside the model's learning process, ensuring robust data protection without sacrificing predictive performance. By integrating AdaLDP within a federated learning framework, FedCCTR-LM not only keeps data localized within its original domain, reducing data breach risks, but also allows nuanced privacy modulation. 

\subsection{Comparison with Existing CCTR Methods in CL Implementation}

Contrastive learning has become a key technique for enhancing cross-domain feature representation in CCTR prediction, with different models adopting diverse strategies to manage domain-specific and cross-domain interactions. Models like C$^{2}$DSR\citep{cao2022c2dsr} primarily use general-purpose CL objectives to align features across domains, effectively enhancing shared representation but often lacking a clear mechanism for balancing domain-specific and cross-domain nuances. C$^{2}$DSR utilizes a global graph that facilitates embedding of both source and target domain features into a unified space. Although these approaches successfully enhance global representation transfer, they fall short in differentiating between domain-specific and cross-domain elements, which can limit their adaptability in mixed-domain scenarios. Tri-CDR\citep{ma2024triple} offers a more nuanced approach by introducing Triple Cross-Domain Attention (TCA) and Triple Contrastive Learning (TCL), focusing on the joint modeling of source, target, and mixed sequences. TCA helps capture domain-specific interactions, while TCL works to align these interactions across domains, enhancing both global user interests and specific domain nuances. However, Tri-CDR does not clearly separate intra-domain and inter-domain objectives, leading to potential limitations in managing mixed-domain personalization effectively. The reliance on complex multi-attention mechanisms also adds computational burden, which may hinder scalability.

In contrast, FedCCTR-LM introduces a refined contrastive learning structure through its IDRA and CDRD modules. IDRA addresses the noise introduced by LLM-based augmentation within each domain, ensuring domain-specific sequences remain coherent and consistent. Meanwhile, CDRD focuses on inter-domain relationships, increasing the separation between domains $A$ and $B$ while drawing both domains closer to the mixed domain. This dual contrastive learning strategy enables FedCCTR-LM to achieve robust cross-domain alignment without sacrificing domain-specific personalization, providing a structured approach that balances both cross-domain generalization and domain-specific accuracy. 

\subsection{Complexity Analysis}

This subsection provides a comprehensive complexity analysis of FedCCTR-LM in comparison with federated CCTR models (FedCTR\citep{wu2022fedctr}, FedDCSR\citep{zhang2024feddcsr}) and the non-federated Tri-CDR\citep{ma2024triple} model. FedCTR, focusing on cross-platform user embedding aggregation, attains the lowest space complexity \(\mathcal{O}(Nd)\) but lacks explicit cross-domain modeling. FedDCSR, employing disentangled representation learning and variational graph-based encoders, has a space complexity of \(\mathcal{O}(L^2 d + 2Ld)\) and a time complexity \(\mathcal{O}(B \cdot L^2 d + B^2 \cdot d)\), driven by batch size \(B\) and sequence length \(L\). Tri-CDR, a centralized cross-domain framework with triple cross-domain attention, incurs \(\mathcal{O}(3Ld + 3L^2)\) in space and \(\mathcal{O}((L^2 + L + B)d|U|)\) in time, posing substantial overhead in large-scale deployments.

In contrast, FedCCTR-LM integrates federated learning with effective cross-domain alignment under a balanced complexity profile. The PrivAugNet module, accessing LLM via API, retains only augmented features locally, contributing \(\mathcal{O}(N d')\) to space, where \(d'\) is the augmented feature dimension. The IDST-CL module adds \(\mathcal{O}(N^2 d)\), and AdaLDP introduces \(\mathcal{O}(d)\) for privacy-preserving noise. Hence, FedCCTR-LM’s total space complexity, \(\mathcal{O}(N d' + N^2 d + d)\), lowers reliance on sequence length, while its time complexity \(\mathcal{O}(B \cdot (L^2d + B \cdot d))\) ensures federated scalability. These characteristics endow FedCCTR-LM with superior efficiency in resource-constrained environments (e.g., mobile devices) relative to FedDCSR and Tri-CDR, making it an applicable choice for practical federated deployments.

\section{Experiments}
\label{sec:experiments}

In this section, we present a comprehensive experimental evaluation to systematically address the following research questions:

\begin{itemize}
    \item \textbf{RQ1}: How does FedCCTR-LM perform compared to state-of-the-art CCTR prediction models? (Section~\ref{subsec:exp_1})
    \item \textbf{RQ2}: What contributions do individual components of the FedCCTR-LM framework make toward its overall performance?  (Section~\ref{subsec:exp_2})
    \item \textbf{RQ3}: How do variations in critical hyperparameters impact the performance of FedCCTR-LM across datasets and configurations?  (Section~\ref{subsec:exp_3})
    \item \textbf{RQ4}: Can the PrivAugNet module be seamlessly integrated into other CTR prediction models to enhance performance? (Section~\ref{subsec:exp_4})
    \item \textbf{RQ5}: How does the AdaLDP, IDRA, and CDRD contribute to the cross-domain representation learning in CCTR? (Section~\ref{subsec:exp_5})
\end{itemize}

\subsection{Datasets}

\begin{table}[width=.9\linewidth,cols=5,pos=h]
\caption{Statistics of the Original and Augmented Datasets}
\label{tab:dataset_stats}
\begin{tabular*}{\tblwidth}{@{} LCC|CC@{} }
\toprule
\textbf{Metric} & \textbf{Book} & \textbf{Movie} & \textbf{Food} & \textbf{Kitchen} \\
\midrule
\# Ori. Feature Fields & User: 0, Item: 1 & User: 0, Item: 1 & User: 0, Item: 1 & User: 0, Item: 1 \\
\# Aug. Feature Fields & User: 3, Item: 4 & User: 3, Item: 4 & User: 3, Item: 4 & User: 3, Item: 4 \\
\# Common Users & \multicolumn{2}{c|}{2,238} & \multicolumn{2}{c}{2,678} \\
\# Unique Items & 28,337 & 55,172 & 23,729 & 34,445 \\
\# Train Insts. & \multicolumn{2}{c|}{614,400} & \multicolumn{2}{c}{491,520} \\
\# Val. Insts. & \multicolumn{2}{c|}{153,600} & \multicolumn{2}{c}{105,300} \\
\# Test Insts. & \multicolumn{2}{c|}{153,600} & \multicolumn{2}{c}{105,300} \\
Ori. Avg Seq. Lens & 13.33 & 5.50 & 12.91 & 4.10 \\
Aug. Avg Seq. Lens & 15.41 & 10.00 & 13.85 & 10.00 \\
Ori. Sparsity & 99.95\% & 99.99\% & 99.95\% & 99.99\% \\
Aug. Sparsity & 99.94\% & 99.98\% & 99.94\% & 99.97\% \\
\bottomrule
\end{tabular*}
\end{table}

\begin{figure*}[!t]
    \centering
    \includegraphics[width=0.8\linewidth, trim={0 0 0 0}, clip]{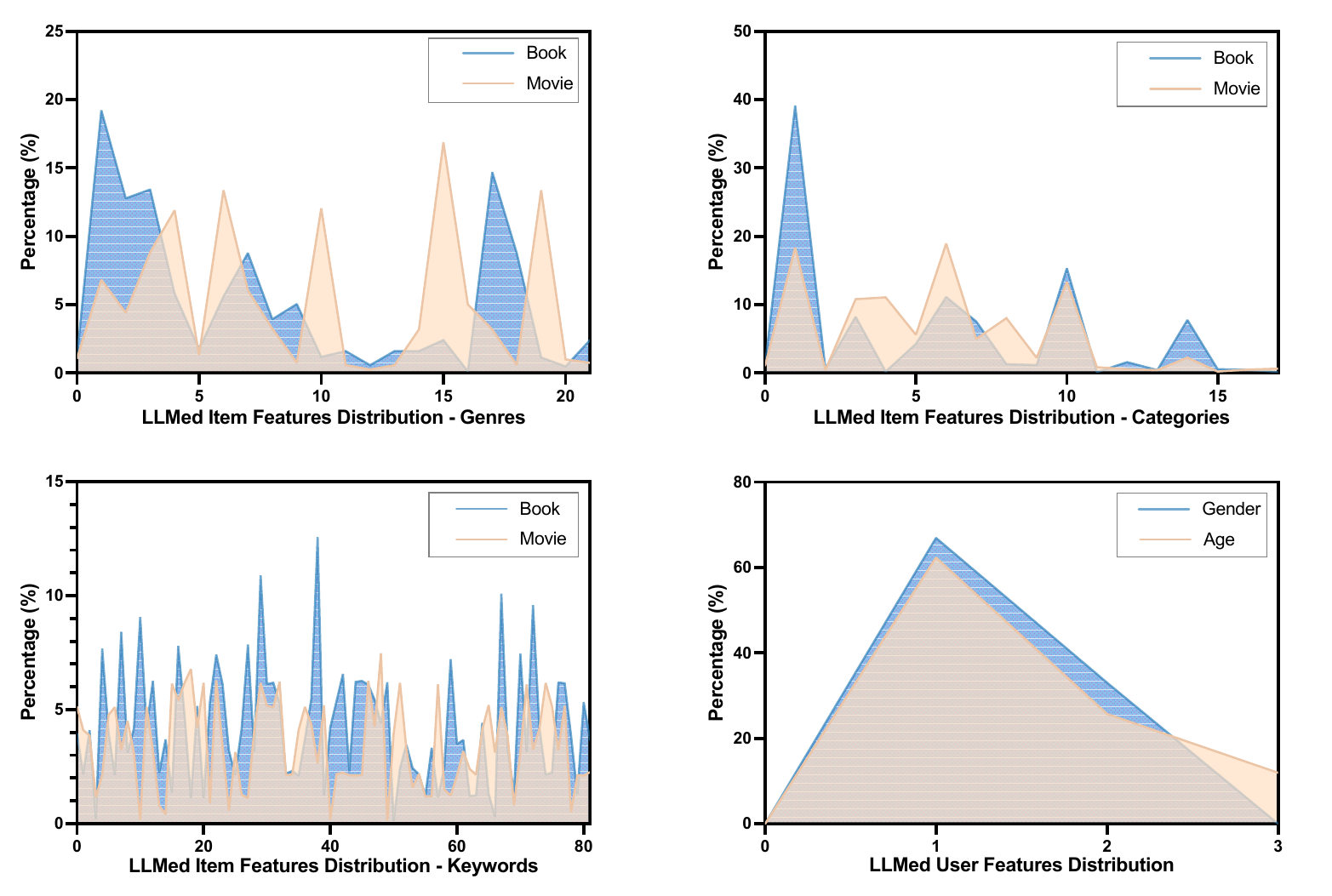}
    \caption{Feature distribution of augmented Book \& Movie datasets using the PrivAugNet module. Enhanced item features include genres, themes, and keywords, while user features include age and gender.}
    \label{fig:feature_distribution}
\end{figure*}

The experiments in this study employ the Amazon dataset\footnote{Amazon products review data: \url{https://cseweb.ucsd.edu/~jmcauley/datasets/amazon/links.html}} to evaluate the effectiveness of the proposed \textbf{FedCCTR-LM} framework. The Amazon dataset is well-suited for CCTR prediction tasks as it includes item interaction data (such as user IDs, item IDs, ratings, and interaction timestamps) from overlapping users across multiple domains. Specifically, two CCTR tasks were constructed: \textit{Book \& Movie} and \textit{Food \& Kitchen}. The \textit{Movie} domain includes users' viewing history, while the \textit{Book} domain includes users' reading history. The \textit{Food} and \textit{Kitchen} domains respectively represent users' food purchase history and kitchenware procurement history. 

The dataset was preprocessed following established methodologies~\citep{li2021dual, li2022recguru}. This process involved selecting users with interactions in both domains and excluding those with insufficient interaction records. Only users with more than 10 interactions and items with occurrence frequencies exceeding 10 were retained. All rating records were interpreted as click behaviors. To construct user behavior sequences for domains $A$, $B$, and the mixed domain $M$, interaction histories were ordered chronologically. A leave-one-out splitting strategy~\citep{geroldinger2023leave} was applied, designating the user’s final interaction as the test set and the penultimate interaction as the validation set.

In addition to the raw data, we applied the proposed \textbf{PrivAugNet} module to enhance the dataset. The augmentation was carried out by utilizing the "LLama-2-13b-chat-hf" conversational model provided by Meta AI through the Hugging Face API~\footnote{\url{https://huggingface.co/docs/transformers/main/en/model_doc/llama2}}. Specifically, for the \textit{Book \& Movie} dataset, item features were enriched to include genres, themes, and keywords, while user features were expanded to include age and gender attributes. The distribution of these enhanced features is illustrated in Figure~\ref{fig:feature_distribution}. The statistical summaries of the original and augmented datasets are provided in Table~\ref{tab:dataset_stats}.

\subsection{Baselines}

To comprehensively evaluate the performance of the proposed FedCCTR-LM framework, we benchmark it against a diverse range of established baseline models. These baselines are categorized into three groups: \textbf{Traditional CTR Models}, \textbf{Cross-Domain CTR Models}, and \textbf{Federated CTR Models}. This classification provide a holistic perspective on CCTR prediction, encompassing conventional approaches, cross-domain techniques, and advanced federated learning strategies prioritizing privacy. All baseline models were meticulously tuned to ensure a fair and consistent comparison under uniform experimental conditions. Below, we detail the models within each category:

\paragraph{Traditional CTR Models:}

\begin{itemize}
    \item[-] \textbf{LR}\citep{agarwal2015comparative}: A foundational linear CTR prediction model serving as a benchmark for evaluating the performance of advanced methods.
    \item[-] \textbf{Item-KNN}\citep{reddy2019content}: A collaborative filtering approach that recommends items based on similarity derived from historical user interactions, adapted here for CTR prediction.
    \item[-] \textbf{DIEN}\citep{zhou2019deep}: An RNN-based model that captures the dynamic evolution of user interests, offering a robust baseline for single-domain CTR prediction tasks.
    \item[-] \textbf{BST}\citep{chen2019behavior}: A transformer-based model designed to capture sequential dependencies within user interaction histories.
\end{itemize}

\paragraph{Cross-Domain CTR Models:}

\begin{itemize}
    \item[-] \textbf{CoNet}\citep{hu2018conet}: A cross-connection architecture facilitating knowledge transfer between domains by jointly modeling user behaviors in both source and target domains. It serves as a key baseline for assessing cross-domain representation learning.
    \item[-] \textbf{MiNet}\citep{ouyang2020minet}: A framework integrating domain-specific and shared representations to model user interests across multiple domains, emphasizing cross-domain learning capabilities.
    \item[-] \textbf{C$^2$DSR}\citep{cao2022c2dsr}: A model leveraging contrastive infomax objectives to enhance global-local correlations across domains, enabling effective cross-domain knowledge transfer through mutual information maximization.
    \item[-] \textbf{Tri-CDR}\citep{ma2024triple}: A framework jointly modeling source, target, and mixed behavior sequences to extract user preferences across domains, effectively enhancing domain-specific knowledge transfer.
\end{itemize}

\paragraph{Federated CTR Models:}

\begin{itemize}
    \item[-] \textbf{FedDSSM}\citep{qin2023split}: An adaptation of the DSSM to federated learning, focusing on preserving user privacy while ensuring effective item recommendations.
    \item[-] \textbf{FedSASRec}: An extension of the SASRec model~\citep{kang2018self} to a federated setting, designed to preserve data privacy while facilitating sequential recommendations.
    \item[-] \textbf{FedCL4SRec}: A federated adaptation of CL4SRec~\citep{xie2022CL4SR}, incorporating contrastive learning to enhance sequential representations.
    \item[-] \textbf{FedDCSR}\citep{zhang2024feddcsr}: A framework employing disentangled representation learning for federated cross-domain recommendation. It incorporates inter-intra domain sequence disentanglement and contrastive infomax objectives to effectively capture both domain-shared and domain-specific features.
\end{itemize}

\subsection{Experimental Settings}

All experiments were conducted on an NVIDIA A100 GPU with 40GB memory, utilizing PyTorch~\citep{paszke2019pytorch} and Python 3.8.10 for implementing the proposed FedCCTR-LM framework and all baseline models. To ensure fair and optimal performance comparisons, an extensive grid search was performed to identify the best hyperparameter configurations for both FedCCTR-LM and the baselines. The AdamW optimizer~\citep{loshchilov2017decoupled} was employed across all models, with a learning rate set to \(5 \times 10^{-4}\) and a batch size of 128. In federated settings, the local batch size for user models was set to 32. All model parameters were initialized using the Xavier initialization method.

Embedding sizes of \(\{16, 32, 64\}\) were evaluated across all models, and the best-performing configuration was selected. For sequence modeling, a maximum sequence length of 20 was uniformly applied to all datasets. The augmentation process used LLM parameters fine-tuned for optimal performance, with temperature values selected from \(\{0, 0.4, 0.8, 1.2\}\) to control randomness in generated text and top-p values chosen from \(\{0, 0.3, 0.6, 0.9\}\). The number of candidate items, \(\mathbf{|C|}\), for interaction sequence expansion was set to \(\{5, 10, 20\}\).

Given the sparsity and variability in data distribution across domains, hyperparameters for the IDST-CL module were adapted to suit different cross-domain datasets. Specifically, \(\lambda_1\) and \(\lambda_2\), which govern the strength of IDRA and CDRD, were selected from the range \([0.0, 1.0]\). The temperature coefficient \(\tau\) was fixed at 0.1. For IDRA, the margin \(\alpha\) was set to 0.5 for denser datasets (e.g., Book and Food) and to 2.0 for sparser datasets (e.g., Movie and Kitchen). The MLP structure used for CTR prediction consisted of three fully connected layers with dimensions \(512 \times 256 \times 128\). The transformer block contained a single layer with a dropout rate of 0.1 and a multi-head attention mechanism employing 8 heads.

For the AdaLDP module, unless explicitly specified, the privacy budget \(\epsilon\) was set to 1, the decay rate \(\mathcal{R}\) to 0.997, and the client sampling ratio \(\rho\) to 0.01. To ensure the reliability and robustness of results, each experiment was repeated five times with different random seeds, and the average performance metrics were reported.

\subsection{Performance Comparison on CCTR Prediction (RQ1)}
\label{subsec:exp_1}

\begin{table}[width=.98\linewidth,cols=14,pos=!t]
    \begin{threeparttable}
    \caption{Performance comparison on Book\&Movie dataset in terms of NDCG@{2,5,10} and MRR@{2,5,10}}
    \label{tab:performance_comparison_B&M}
    \setlength{\tabcolsep}{5.5pt}
        \begin{tabular*}{\tblwidth}{@{}LLCCCCCCCCCCCC@{}}
        \toprule
        & & \multicolumn{6}{c}{\textbf{Book}} & \multicolumn{6}{c}{\textbf{Movie}} \\
        \cmidrule(lr){3-8} \cmidrule(lr){9-14}
        \textbf{Model} & \textbf{Metrics} & \multicolumn{2}{c}{\textbf{K=2}} & \multicolumn{2}{c}{\textbf{K=5}} & \multicolumn{2}{c}{\textbf{K=10}} & \multicolumn{2}{c}{\textbf{K=2}} & \multicolumn{2}{c}{\textbf{K=5}} & \multicolumn{2}{c}{\textbf{K=10}} \\
        & & NDCG & MRR & NDCG & MRR & NDCG & MRR & NDCG & MRR & NDCG & MRR & NDCG & MRR \\
        \midrule
        \multirow{4}{*}{Traditional} & LR & 3.10 & 0.08 & 5.50 & 0.12 & 6.69 & 0.14 & 2.50 & 0.06 & 4.16 & 0.09 & 4.89 & 0.11 \\
        & Item-KNN & 5.77 & 1.05 & 6.56 & 1.21 & 8.51 & 1.95 & 4.58 & 0.85 & 7.55 & 1.15 & 8.87 & 1.57 \\
        & DIEN & 15.03 & 9.20 & 16.56 & 9.94 & 17.14 & 10.21 & 13.09 & 7.04 & 14.49 & 7.73 & 15.41 & 8.51 \\
        & BST & 18.52 &	12.48 &	20.07 &	12.85 &	20.97 &	13.89 &	14.42 &	10.27 &	15.75&	11.08 &	16.51 &	11.47 \\
        \midrule
        \multirow{4}{*}{Cross-Domain} & CoNet & 6.05 & 2.43 & 7.45 & 2.77 & 8.75 & 3.32 & 5.51 & 2.15 & 6.87 & 2.70 & 7.90 & 3.05 \\
        & MiNet & 13.55 & 7.52 & 15.25 & 8.21 & 16.52 & 8.45 & 13.20 & 7.51 & 14.72 & 8.65 & 17.36 & 8.72 \\
        & C$^2$DSR & \underline{19.10} & \underline{13.22} & \underline{20.25} & \underline{14.05} & \underline{21.95} & \underline{15.33} & 16.75 & \underline{12.79} & 18.13 & 13.15 & 19.42 & 13.86 \\
        & Tri-CDR & 18.19 & 12.81 & 19.75 & 13.55 & 20.78 & 14.43 & \underline{17.56} & 12.52 & \underline{19.05} & \underline{13.31} & \underline{20.09} & \underline{13.88} \\
        \midrule
        \multirow{4}{*}{Federated} & FedDSSM & 6.57 & 4.57 & 8.28 & 5.05 & 8.75 & 5.47 & 5.80 & 3.77 & 7.06 & 4.50 & 8.14 & 4.85 \\
        & FedSASRec & 16.75 & 11.50 & 18.20 & 12.05 & 19.10 & 13.05 & 13.86 & 8.15 & 14.38 & 9.10 & 16.18 & 11.45 \\
        & FedCL4SRec & 18.06 & 12.63 & 19.59 & 13.35 & 20.21 & 14.95 & 15.12 & 11.58 & 16.32 & 12.35 & 17.79 & 12.70 \\
        & FedDCSR & 18.10 & 12.70 & 19.60 & 13.40 & 20.15 & 14.79 & 15.15 & 11.57 & 16.39 & 12.32 & 17.89 & 12.75 \\
        \midrule
        \multirow{2}{*}{Our Model} & FedCCTR-LM & \textbf{20.12} & \textbf{14.07} & \textbf{21.65} & \textbf{14.62} & \textbf{23.10} & \textbf{15.71} & \textbf{19.15} & \textbf{13.62} & \textbf{20.79} & \textbf{14.21} & \textbf{22.19} & \textbf{14.74} \\
        \cmidrule(lr){2-14}
        & \textbf{Improv.} & 5.34\% & 6.43\% & 6.91\% & 4.06\% & 5.24\% & 2.48\% & 9.05\% & 6.49\% & 9.13\% & 6.76\% & 10.45\% & 6.20\% \\
        \bottomrule
    \end{tabular*}
    \begin{tablenotes}
        \footnotesize
        \item \textbf{Note:} The highest performance in each column is shown in bold, while the second-highest performance is underlined. All improvements (Improv.) are statistically significant based on a paired t-test with $p < 0.05$.
    \end{tablenotes}
    \end{threeparttable}
\end{table}

\begin{table}[width=.98\linewidth,cols=14,pos=!h]
    \centering
    \begin{threeparttable}
    \caption{Performance comparison on Food\&Kitchen dataset in terms of NDCG@{2,5,10} and MRR@{2,5,10}}
    \label{tab:performance_comparison_F&K}
    \setlength{\tabcolsep}{5.5pt}
    \begin{tabular*}{\tblwidth}{@{}LLCCCCCCCCCCCC@{}}
        \toprule
        & & \multicolumn{6}{c}{\textbf{Food}} & \multicolumn{6}{c}{\textbf{Kitchen}} \\
        \cmidrule(lr){3-8} \cmidrule(lr){9-14}
        \textbf{Model} & \textbf{Metrics} & \multicolumn{2}{c}{\textbf{K=2}} & \multicolumn{2}{c}{\textbf{K=5}} & \multicolumn{2}{c}{\textbf{K=10}} & \multicolumn{2}{c}{\textbf{K=2}} & \multicolumn{2}{c}{\textbf{K=5}} & \multicolumn{2}{c}{\textbf{K=10}} \\
        & & NDCG & MRR & NDCG & MRR & NDCG & MRR & NDCG & MRR & NDCG & MRR & NDCG & MRR \\
        \midrule
        \multirow{4}{*}{Traditional} & LR & 5.67 & 1.04 & 8.11 & 1.11 & 9.98 & 0.67 & 2.83 & 0.06 & 5.20 & 0.09 & 7.75 & 0.24 \\
        & Item-KNN & 9.57 & 2.01 & 14.42 & 2.60 & 17.05 & 2.96 & 4.97 & 1.06 & 6.79 & 1.42 & 8.88 & 1.90 \\
        & DIEN & 19.78 & 13.84 & 22.47 & 14.68 & 23.21 & 15.05 & 13.66 & 7.59 & 15.35 & 8.33 & 16.81 & 9.29 \\
        & BST & 23.58 & 15.58 & 24.93 & 16.12 & 26.15 & 17.07 & 16.29 & 11.77 & 18.28 & 12.67 & 20.38 & 12.26 \\
        \midrule
        \multirow{4}{*}{Cross-Domain} & CoNet & 9.74 & 4.49 & 11.48 & 5.47 & 13.61 & 6.98 & 7.25 & 2.87 & 8.73 & 3.37 & 10.33 & 4.00 \\
        & MiNet & 18.19 & 9.78 & 20.08 & 10.38 & 22.01 & 11.05 & 15.74 & 8.68 & 17.12 & 9.34 & 19.30 & 10.10 \\
        & C$^2$DSR & \underline{24.46} & 17.28 & \underline{26.32} & \underline{17.89} & \underline{28.28} & \underline{18.41} & 18.84 &	14.17 & 21.22 & 14.88 & 22.31 & 15.61  \\
        & Tri-CDR & 22.13 & 15.91 & 24.89 & 16.59 & 26.35 & 17.81 & \underline{21.14} & \underline{14.41} & \underline{22.79} & \underline{15.97} & \underline{24.80} & \underline{16.85} \\
        \midrule
        \multirow{4}{*}{Federated} & FedDSSM & 9.58 & 5.68 & 11.33 & 6.08 & 11.63 & 6.89 & 7.57 & 5.17 & 9.07 & 5.73 & 10.11 & 6.33 \\
        & FedSASRec & 21.38 & 15.55 & 22.90 & 16.37 & 24.36 & 16.88 & 17.37 & 11.30 & 18.28 & 12.87 & 18.34 & 11.95 \\
        & FedCL4SRec & 22.01 & 15.94 & 23.38 & 16.65 & 25.60 & 17.73 & 18.01 & 11.87 & 19.18 & 13.97 & 20.69 & 14.83 \\
        & FedDCSR & 23.04 & \underline{17.30} & 24.20 & 17.63 & 26.34 & 18.28 & 17.83 & 12.03 & 18.99 & 13.65 & 20.70 & 14.82 \\
        \midrule
        \multirow{2}{*}{Our Model} & FedCCTR-LM & \textbf{25.76} & \textbf{17.67} & \textbf{27.69} & \textbf{18.41} & \textbf{29.23} & \textbf{19.67} & \textbf{23.08} & \textbf{16.40} & \textbf{24.81} & \textbf{17.09} & \textbf{26.68} & \textbf{17.92} \\
        \cmidrule(lr){2-14}
        & \textbf{Improv.} & 5.31\% & 2.12\% & 5.20\% & 2.90\% & 3.34\% & 6.84\% & 9.18\% & 13.83\% & 8.85\% & 7.05\% & 7.59\% & 6.31\% \\
        \bottomrule
    \end{tabular*}
    \begin{tablenotes}
        \footnotesize
        \item \textbf{Note:} The highest performance in each column is shown in bold, while the second-highest performance is underlined. All improvements (Improv.) are statistically significant based on a paired t-test with $p < 0.05$.
    \end{tablenotes}
    \end{threeparttable}
\end{table}

This section evaluates the comparative performance of the proposed FedCCTR-LM framework against traditional, cross-domain, and federated CTR models. Metrics include NDCG and MRR at cutoff values \(K = \{2, 5, 10\}\), which assess ranking relevance and the position of the first relevant item, respectively. To ensure robustness, each positive instance was paired with 99 randomly sampled negatives during testing. The performance results for the \textit{Book \& Movie} and \textit{Food \& Kitchen} datasets are summarized in Tables~\ref{tab:performance_comparison_B&M} and~\ref{tab:performance_comparison_F&K}.

Across all datasets, FedCCTR-LM consistently outperformed baseline models, achieving substantial gains in all metrics. For instance, in NDCG@10, it recorded improvements of 5.24\%, 10.45\%, 3.34\%, and 7.59\% over the best-performing baseline models in the Book, Movie, Food, and Kitchen datasets, respectively. Similarly, in MRR@10, it demonstrated enhancements of 2.48\%, 6.20\%, 6.84\%, and 6.31\%. These results underscore the synergistic contributions of PrivAugNet, IDST-CL, and AdaLDP, which collectively address challenges like data sparsity, feature incompleteness, cross-domain knowledge transfer, and the privacy-utility trade-off.

FedCCTR-LM showed substantial advantages compare to traditional CTR models, especially in sparse datasets like \textit{Movie} and \textit{Kitchen}. For example, in NDCG@10, it achieved scores of 22.19 and 26.68, outperforming DIEN by 43.99\% and 58.71\%, and BST by 34.40\% and 30.91\%. While DIEN and BST are adept at capturing sequential patterns in single-domain contexts, they lack the capability to handle cross-domain sparsity and heterogeneity effectively. 

Also, FedCCTR-LM demonstrated clear superiority against CCTR models. For example, it achieved an NDCG@10 score of 23.10 on the Book dataset, surpassing C$^2$DSR by 5.24\%, and improved by 10.45\% over Tri-CDR on the Movie dataset. These improvements highlight the role of PrivAugNet in enriching sparse datasets and the effectiveness of IDST-CL in aligning domain-specific and shared representations. 

Finally, in federated CTR settings, FedCCTR-LM demonstrated significant gains over models like FedSASRec, FedCL4SRec, and FedDCSR. On the Book dataset, it improved NDCG@10 by 14.64\% compared to FedDCSR, while on the Kitchen dataset, it achieved a 6.95\% higher MRR@10 than FedCL4SRec. These results highlight the robustness of FedCCTR-LM’s integrated design in addressing federated learning challenges, particularly Non-I.I.D. data distributions and privacy-utility trade-offs, achieving superior generalization and privacy-preserving performance.

\subsection{Ablation Study (RQ2)}
\label{subsec:exp_2}

\begin{figure*}[!t]
    \centering
    \includegraphics[width=\linewidth, trim={0 0.2cm 0 0}, clip]{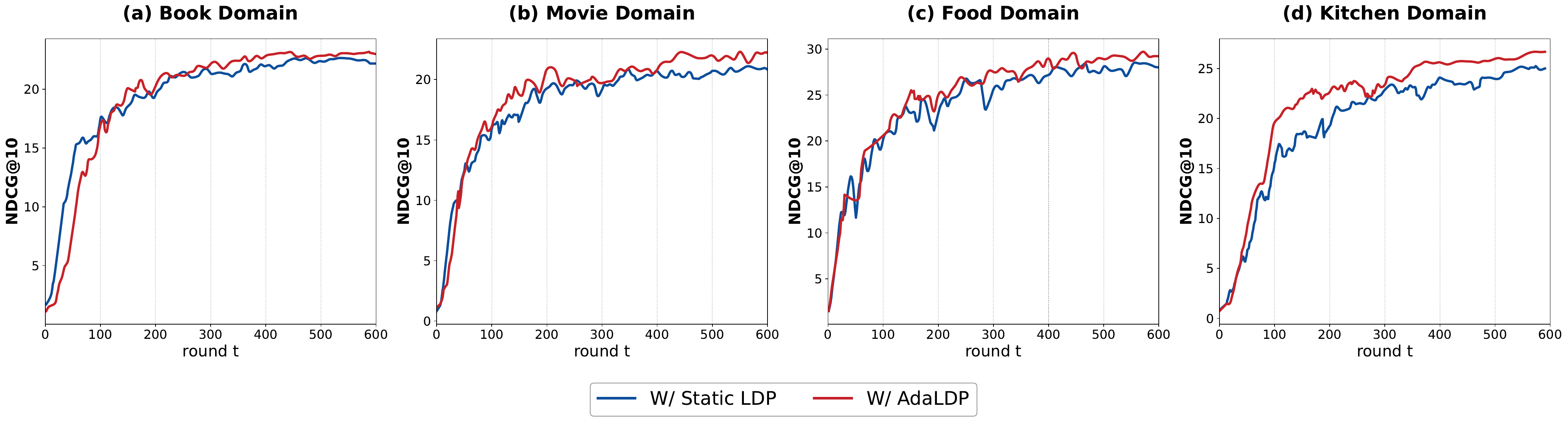}
    \caption{The Comparison of NDCG@10 Performance and Convergence Speed between FedCCTR-LM with AdaLDP and with Static LDP on Book \& Movie, Food \& Kitchen Datasets.}
    \label{fig:impact_AdaLDP}
\end{figure*}

\begin{table}[width=.9\linewidth,cols=9,pos=!t]
\centering
\caption{Ablation Study on the Performance of FedCCTR-LM Across Different Variants in Terms of NDCG@10 and MRR@10 on Book \& Movie, Food \& Kitchen Datasets.}
\label{tab:ablation_study}
\begin{threeparttable}
\setlength{\tabcolsep}{6pt} 
\renewcommand{\arraystretch}{1.2} 
\begin{tabular*}{\tblwidth}{@{}LCCCCCCCC@{}}
\toprule
\multirow{2}{*}{\textbf{Variants}} & \multicolumn{4}{c}{\textbf{Book}} & \multicolumn{4}{c}{\textbf{Movie}} \\
\cmidrule(lr){2-5} \cmidrule(lr){6-9}
& \textbf{NDCG@10} & \textbf{Diff.} & \textbf{MRR@10} & \textbf{Diff.} 
& \textbf{NDCG@10} & \textbf{Diff.} & \textbf{MRR@10} & \textbf{Diff.} \\
\midrule
w/o PrivAugNet & 21.51 & -7.39\% & 14.77 & -6.36\% & 20.13 & -10.23\% & 13.55 & -8.78\% \\
w/o IDRA       & 22.08 & -4.62\% & 15.24 & -3.08\% & 21.00 & -5.67\% & 14.21 & -3.73\% \\
w/o CDRD       & 21.93 & -5.34\% & 15.20 & -3.36\% & 20.90 & -6.17\% & 14.17 & -4.02\% \\
w/o AdaLDP     & 23.97 & 3.63\%  & 16.25 & 3.32\%  & 23.21 & 4.39\%  & 15.48 & 4.78\% \\
w/ Static LDP  & 22.19 & -4.10\% & 15.09 & -4.11\% & 20.78 & -6.79\% & 13.90 & -6.04\% \\
\midrule
\textbf{FedCCTR-LM} & \textbf{23.10} & - & \textbf{15.71} & - & \textbf{22.19} & - & \textbf{14.74} & - \\
\midrule
\multirow{2}{*}{\textbf{}} & \multicolumn{4}{c}{\textbf{Food}} & \multicolumn{4}{c}{\textbf{Kitchen}} \\
\cmidrule(lr){2-5} \cmidrule(lr){6-9}
& \textbf{NDCG@10} & \textbf{Diff.} & \textbf{MRR@10} & \textbf{Diff.} 
& \textbf{NDCG@10} & \textbf{Diff.} & \textbf{MRR@10} & \textbf{Diff.} \\
\midrule
w/o PrivAugNet & 27.77 & -5.26\% & 18.69 & -5.24\% & 24.01 & -11.12\% & 16.12 & -11.17\% \\
w/o IDRA       & 28.35 & -3.10\% & 19.08 & -3.09\% & 24.69 & -8.06\% & 16.99 & -5.47\% \\
w/o CDRD       & 27.76 & -5.30\% & 19.00 & -3.53\% & 24.48 & -8.99\% & 16.89 & -6.10\% \\
w/o AdaLDP     & 30.12 & 2.95\%  & 20.20 & 2.65\%  & 27.90 & 4.37\%  & 18.57 & 3.50\% \\
w/ Static LDP  & 28.00 & -4.39\% & 18.92 & -3.96\% & 25.01 & -6.68\% & 17.09 & -4.86\% \\
\midrule
\textbf{FedCCTR-LM} & \textbf{29.23} & - & \textbf{19.67} & - & \textbf{26.68} & - & \textbf{17.92} & - \\
\bottomrule
\end{tabular*}
\begin{tablenotes}
    \footnotesize
    \item \textbf{Note:} \textbf{w/o} denotes "without," indicating that the specified module is excluded from the model, while \textbf{w/} denotes "with," indicating inclusion of the specified configuration. FedCCTR-LM represents the baseline model with all modules included. All \textbf{Diff.} values represent the percentage change in performance relative to FedCCTR-LM.
\end{tablenotes}
\end{threeparttable}
\end{table}

To systematically evaluate the contributions of individual components in FedCCTR-LM, we conducted an ablation study by selectively disabling key modules—PrivAugNet, IDRA, CDRD, and AdaLDP—and by substituting AdaLDP with a static differential privacy mechanism (Static LDP). The results, detailed in Table~\ref{tab:ablation_study} and Figure~\ref{fig:impact_AdaLDP}, quantify the impact of these variations on NDCG@10 and MRR@10 across the four datasets, providing critical insights into the functionality of each module.

\textbf{Effect of PrivAugNet} FedCCTR-LM w/o PrivAugNet results in marked performance declines, especially in sparse domains, with NDCG@10 reductions of 7.39\% on Book and 10.23\% on Movie domain, and MRR@10 drops of 6.36\% and 8.78\%, respectively. This highlights PrivAugNet’s importance in enriching data representations, particularly where data density is low, by augmenting user and item features through LLMs to mitigate sparsity and feature incompleteness.

\textbf{Effect of IDRA} Disabling IDRA leads to moderate but consistent declines in performance, with NDCG@10 reductions of 4.62\% (Book), 5.67\% (Movie), 3.10\% (Food), and 5.47\% (Kitchen). MRR@10 exhibits a similar pattern, with decreases of up to 5.47\%. These findings highlight IDRA’s critical function in aligning augmented representations with authentic user preferences, reducing noise introduced during augmentation. By ensuring that domain-specific preferences are preserved, IDRA enhances the model’s ability to capture fine-grained interaction patterns.

\textbf{Effect of CDRD} The absence of CDRD disproportionately impacts sparse datasets, with NDCG@10 declining by 6.17\% on the Movie domain and 6.10\% on the Kitchen domain. MRR@10 similarly drops by 4.02\% and 6.10\%, respectively. This result demonstrates CDRD’s essential role in disentangling domain-specific and cross-domain representations, mitigating negative transfer, and preserving domain uniqueness. By isolating shared and distinct features, CDRD prevents interference across domains, enabling effective knowledge transfer.

\textbf{Effect of AdaLDP} Replacing AdaLDP with Static LDP reveals the advantages of adaptive noise modulation. As illustrated in Figure~\ref{fig:impact_AdaLDP}, FedCCTR-LM with AdaLDP initially converges more slowly due to higher noise levels during early training, reflecting stronger privacy protection. However, around the 200th training round, adaptive noise scaling facilitates more effective gradient updates, leading to faster convergence and higher final NDCG@10 scores across all datasets. In contrast, Static LDP exhibits suboptimal performance, with NDCG@10 declining by 4.10\% to 6.79\% across domains due to its inability to adjust noise dynamically. Removing AdaLDP entirely (w/o AdaLDP) results in marginally better performance but compromises privacy guarantees, emphasizing the necessity of AdaLDP for achieving an optimal balance between privacy preservation and predictive accuracy.

The ablation results collectively validate that the synergy among PrivAugNet, IDRA, CDRD, and AdaLDP is integral to FedCCTR-LM’s effectiveness. PrivAugNet addresses sparsity and feature incompleteness, while IDRA and CDRD ensure robust representation learning by maintaining domain-specific integrity and enabling cross-domain alignment. AdaLDP, with its dynamic noise scaling, strikes a critical balance between privacy and utility, distinguishing FedCCTR-LM as a comprehensive solution for federated cross-domain CTR prediction.

\subsection{Parameter Sensitivity Analyses (RQ3)}
\label{subsec:exp_3}

\begin{figure*}[!t]
    \centering
    \includegraphics[width=\linewidth, trim={0 0.2cm 0 0}, clip]{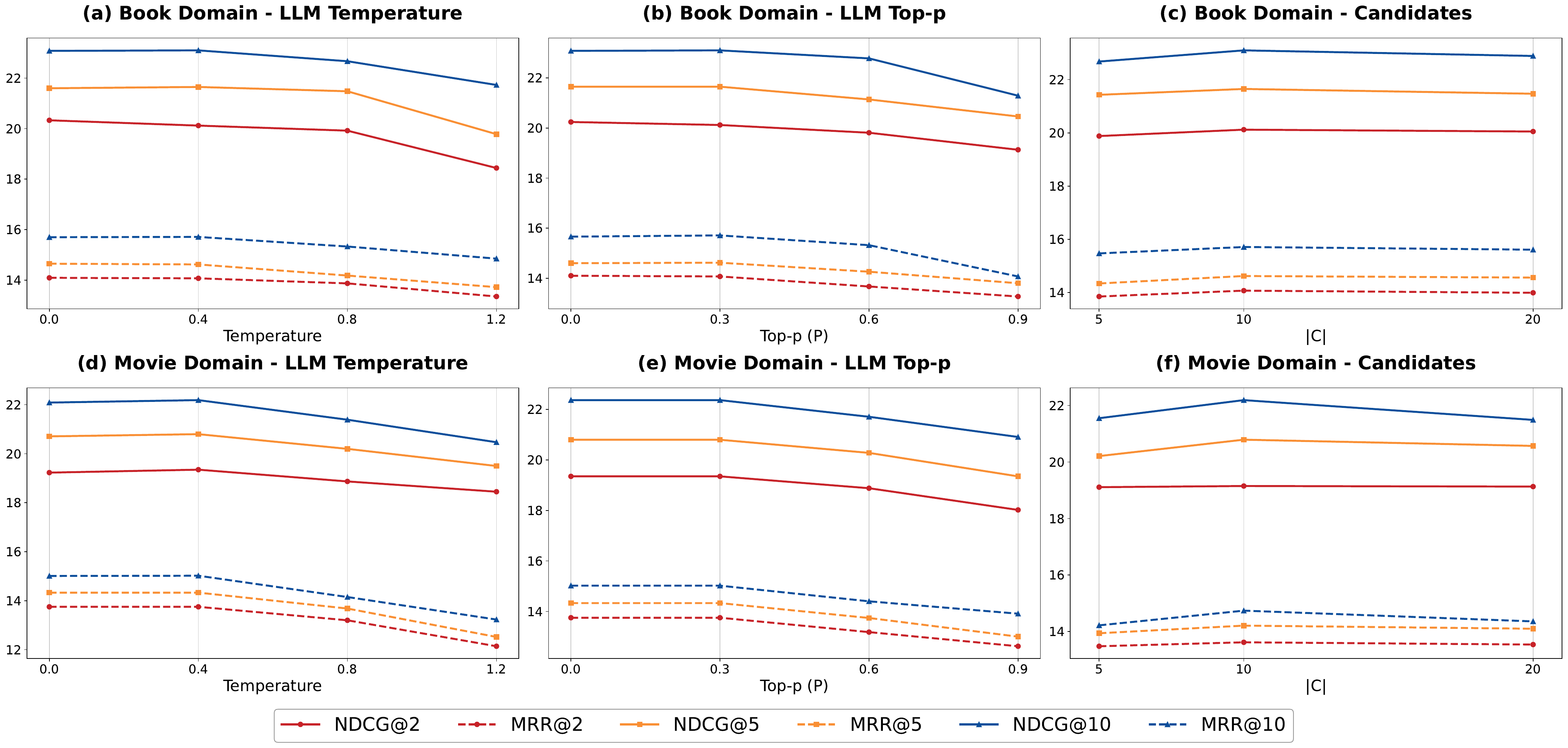}
    \caption{Impact of Key Hyper-parameters in PrivAugNet on Model Performance Across Book and Movie Datasets: Temperature, Top-p Sampling (\(P\)), and Candidate Item Size (\(|\mathcal{C}|\))}
    \label{fig:HP_PrivAugNet}
\end{figure*}
\begin{figure*}[!ht]
    \centering
    \includegraphics[width=\linewidth, trim={0 0.2cm 0 0}, clip]{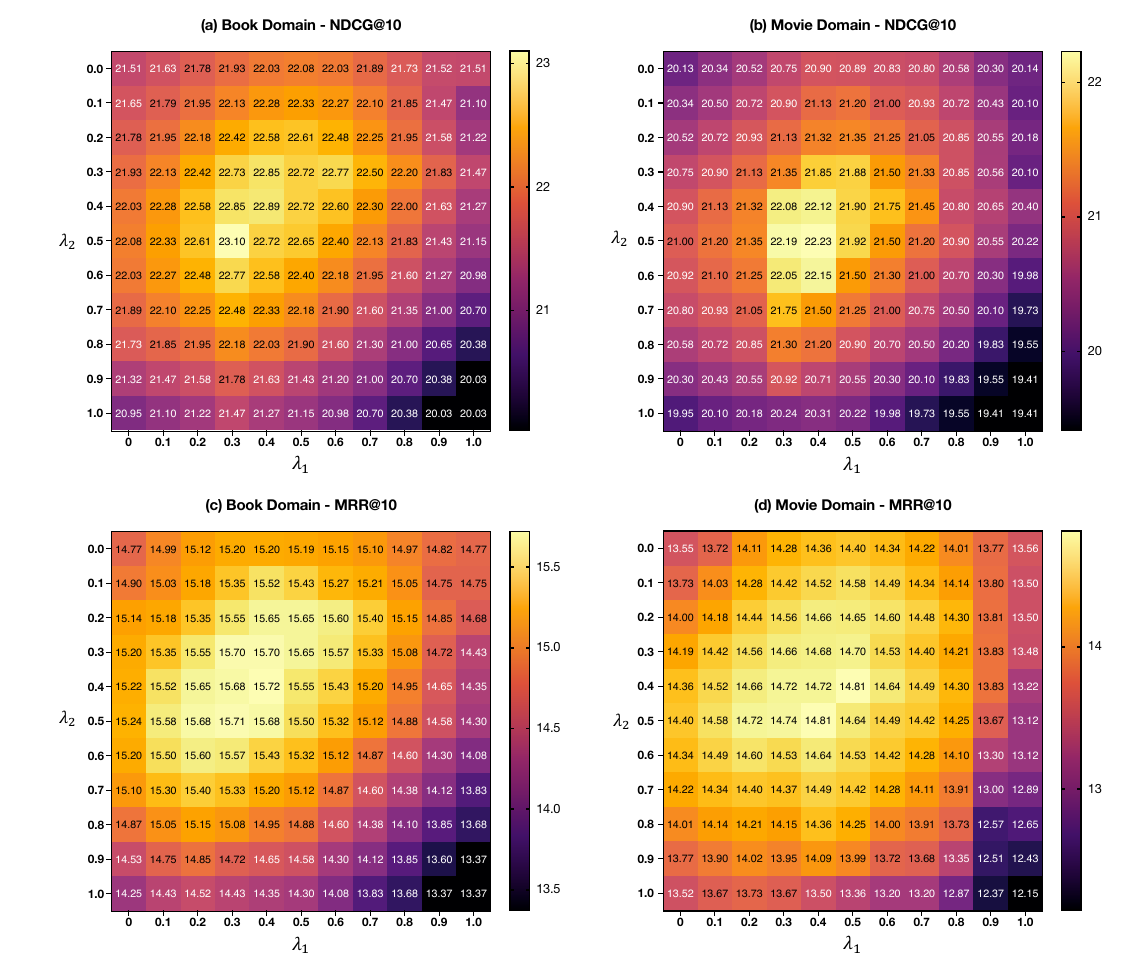}
    \caption{Sensitivity Analysis of Hyperparameters \(\lambda_1\) and \(\lambda_2\) on the Performance of FedCCTR-LM in Terms of NDCG@10 and MRR@10 across Book and Movie Datasets}
    \label{fig:HP_IDST-CL}
\end{figure*}
\begin{figure*}[!t]
    \centering
    \includegraphics[width=\linewidth, trim={0 0.25cm 0 0}, clip]{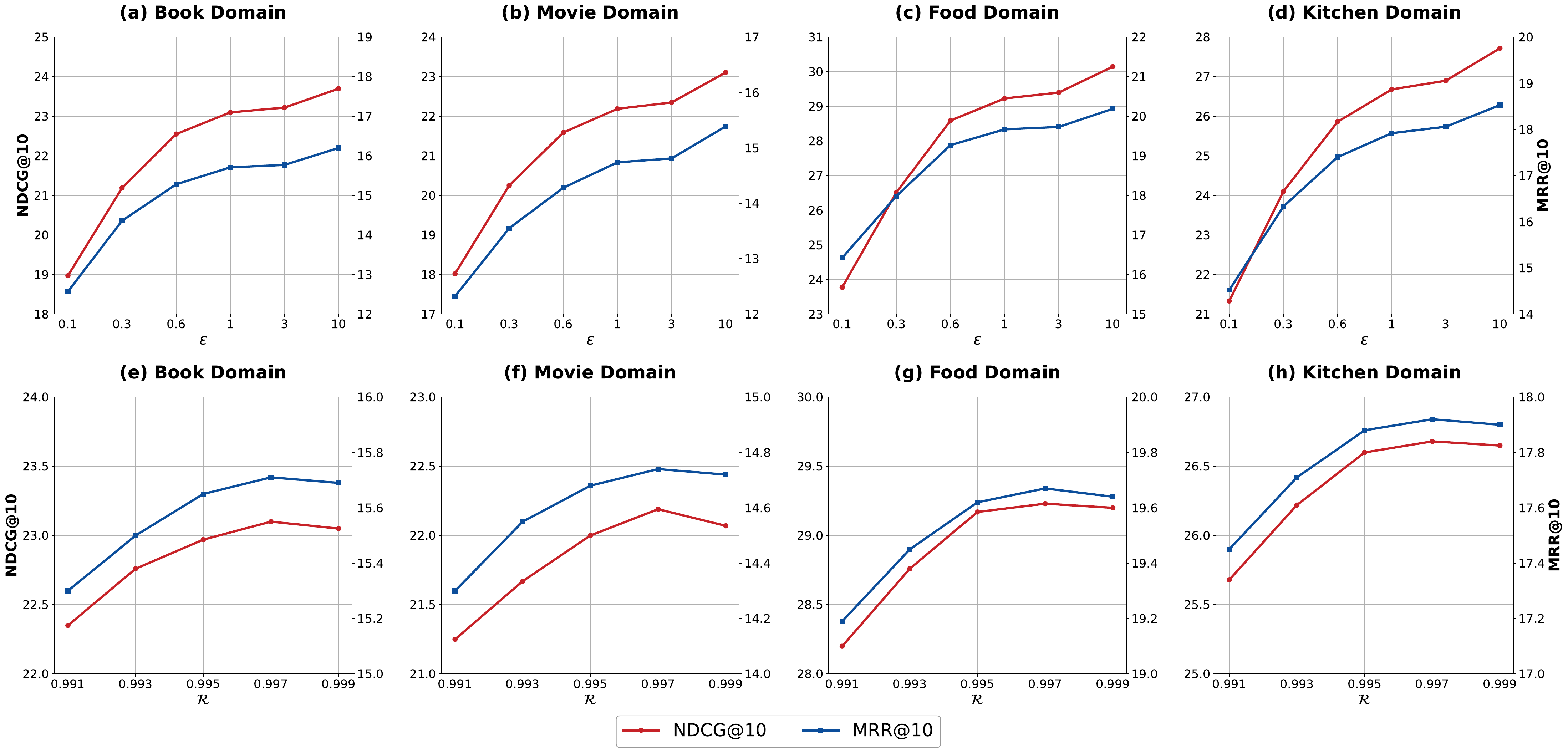}
    \caption{Impact of Privacy Budget \(\epsilon\) and Decay Rate \(\mathcal{R}\) in AdaLDP on Model Performance Across Book\&Movie, Food\&Kitchen Datasets.}
    \label{fig:HP-AdaLDP}
\end{figure*}

We first conducted a comprehensive sensitivity analysis of three critical hyperparameters in the \textbf{PrivAugNet} module: LLM Temperature (temp), Top-p Sampling \((P)\), and the number of Candidate Items \((|\mathcal{C}|)\). The experiments, performed on the Book and Movie datasets, reveal nuanced relationships between these parameters and the model’s performance, as illustrated in Figure~\ref{fig:HP_PrivAugNet}.

\subsubsection{Impact of LLM Temperature} The temperature parameter controls the stochasticity of the LLM’s output, influencing feature diversity. As shown in Figure~\ref{fig:HP_PrivAugNet} (a) and (d), performance peaks at \(\textit{temp} = 0.4\), yielding the highest NDCG@10 scores of 23.10 for the Book dataset and 22.19 for the Movie dataset. Lower temperatures (e.g., \(\textit{temp} = 0\)) result in overly deterministic outputs, limiting diversity, while higher temperatures (e.g., \(\textit{temp} \geq 0.8\)) introduce excessive variability, potentially generating noisy or irrelevant features. The observed decline in performance beyond \(\textit{temp} = 0.8\) suggests that maintaining a balance between diversity and relevance is critical for optimizing user-item representations.

\subsubsection{Impact of LLM Top-p Sampling \((P)\)} Top-p sampling determines the cutoff threshold for token selection, influencing the diversity of augmented content. Figures~\ref{fig:HP_PrivAugNet} (b) and (e) show that performance is optimal when \(p\) is set between 0.3 and 0.6, achieving NDCG@10 scores of 23.10 and 22.37 on the Book and Movie datasets, respectively. Lower values (e.g., \(p = 0\)) restrict token selection to highly probable options, limiting augmentation diversity. Conversely, higher values (e.g., \(p = 0.9\)) include less relevant tokens, diluting the quality of the generated features and leading to a performance decline. This trend underscores the importance of calibrating \(p\) to ensure diverse yet semantically coherent augmentations.

\subsubsection{Impact of Candidate Item Size \((|\mathcal{C}|)\)} The \((|\mathcal{C}|)\) parameter controls the number of items provided in \(\mathcal{P}_\text{seq}\) during the LLM-based sequence expansion phase. A smaller candidate set (e.g., \(|\mathcal{C}| = 5\)) limits the diversity of items, potentially constraining the model's ability to enrich user interactions. Conversely, a larger candidate set (e.g., \(|\mathcal{C}| = 20\)) can introduce irrelevant noise, reducing the overall relevance of the generated features. As illustrated in Figure~\ref{fig:HP_PrivAugNet} (c) and (f), our analysis shows that the optimal candidate size is \(|\mathcal{C}| = 10\), yielding the highest NDCG@10 and MRR@10 across both datasets, with scores of 23.10 and 22.19 for the Book and Movie datasets, respectively. Increasing \(|\mathcal{C}|\) beyond this point leads to a decline in performance, particularly on the higher sparsity Movie dataset.

\subsubsection{Impact of CL Loss Weights \(\lambda_1\) and \(\lambda_2\)}

We further conducted a detailed sensitivity analysis of the hyperparameters \(\lambda_1\) and \(\lambda_2\), which respectively control the weights of the IDRA and CDRD loss terms in the IDST-CL module. These parameters are pivotal for balancing intra-domain alignment and cross-domain disentanglement. Experiments on the Book and Movie datasets reveal nuanced interactions between these hyperparameters, as shown in Figure~\ref{fig:HP_IDST-CL}. NDCG@10 and MRR@10 scores were evaluated across \(\lambda_1\) and \(\lambda_2\) values ranging from 0.0 to 1.0 with a step size of 0.1.

\(\lambda_1\) controls the alignment of augmented sequences with original interaction sequences, reducing augmentation noise. Figures~\ref{fig:HP_IDST-CL}(a) and (c) show optimal \(\lambda_1\) values of 0.3 (Book) and 0.4 (Movie), yielding peak NDCG@10 scores of 23.10 and 22.23, respectively. Performance declines sharply for \(\lambda_1 > 0.6\), as excessive alignment constraints reduce representational flexibility, especially in sparse datasets. Conversely, \(\lambda_1 < 0.2\) under-regularizes augmented features, diminishing intra-domain consistency.

\(\lambda_2\) balances shared and domain-specific representations. Optimal values of \(\lambda_2 = 0.5\) (Figures~\ref{fig:HP_IDST-CL}(b) and (d)) achieve the highest NDCG@10 scores for both datasets (23.10 for Book, 22.23 for Movie). Larger \(\lambda_2\) values (e.g., \(\lambda_2 > 0.7\)) overemphasize disentanglement, fragmenting shared features and degrading generalization. Smaller values (\(\lambda_2 < 0.3\)) fail to sufficiently differentiate shared and domain-specific representations, limiting cross-domain transfer.

The best results occur when \(\lambda_1\) and \(\lambda_2\) are balanced (e.g., \(\lambda_1 = 0.3\), \(\lambda_2 = 0.5\) for Book; \(\lambda_1 = 0.4\), \(\lambda_2 = 0.5\) for Movie). Overemphasizing either parameter (\(\lambda_1, \lambda_2 > 0.6\)) over-regularizes the model, reducing adaptability. Conversely, very low values (\(\lambda_1, \lambda_2 < 0.2\)) underutilize alignment and disentanglement mechanisms, leading to suboptimal performance.

\subsubsection{Impact of Privacy Budget (\(\epsilon\))}

The privacy budget \(\epsilon\) directly influences the trade-off between privacy protection and model accuracy in FedCCTR-LM. As depicted in Figure~\ref{fig:HP-AdaLDP} (a)-(d), when \(\epsilon\) is small (e.g., \(\epsilon = 0.1\)), the added noise severely disrupts the gradient information, leading to significant performance degradation across all datasets. As the privacy budget increases, the noise level decreases, allowing the model to capture user preferences more effectively. This trend continues up to \(\epsilon = 3.0\), where NDCG@10 and MRR@10 stabilize, indicating diminishing returns beyond this point. A moderate privacy budget (e.g., \(\epsilon \in [1, 3]\)) offers a good balance between privacy preservation and prediction accuracy, as it reduces noise without overly compromising model utility. Further increasing \(\epsilon\) to 10 provides minimal performance gains, suggesting an optimal range for practical applications.

\subsubsection{Impact of Decay Rate (\(\mathcal{R}\))}

The noise decay rate \(\mathcal{R}\) in FedCCTR-LM governs the rate at which noise diminishes during training. Figure~\ref{fig:HP-AdaLDP} (e)-(h) shows that a higher decay rate (e.g., \(\mathcal{R} = 0.997\)) consistently yields the best results across all datasets. This setting allows the noise to gradually decrease, enabling effective learning while maintaining sufficient privacy in the initial training phases. For example, at \(\mathcal{R} = 0.997\), NDCG@10 for the Movie dataset reaches 22.19. A lower decay rate (e.g., \(\mathcal{R} = 0.991\)) introduces excessive noise early on, hindering the model's convergence. On the other hand, a very high decay rate (e.g., \(\mathcal{R} = 0.999\)) reduces noise too rapidly, potentially exposing sensitive data and leading to overfitting. Thus, selecting an intermediate decay rate, such as \(\mathcal{R} = 0.996\), balances privacy and performance effectively.

\subsection{Model-agnostic Property of PrivAugNet (RQ4)}
\label{subsec:exp_4}

\begin{figure*}[!t]
    \centering
    \includegraphics[width=0.9\linewidth, trim={0 0.2cm 0 0}, clip]{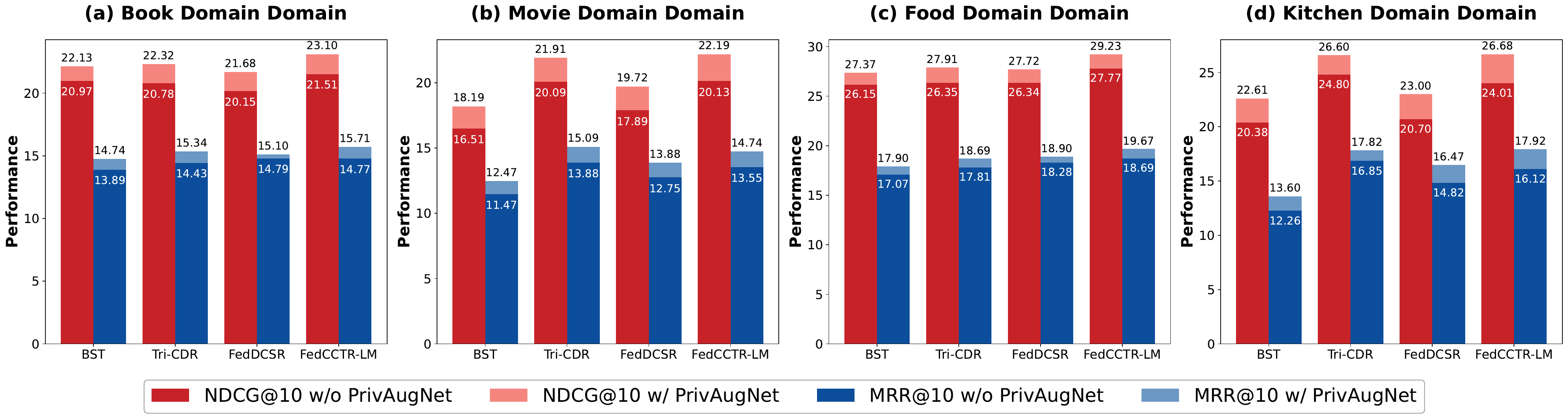}
    \caption{Evaluating the Model-Agnostic of PrivAugNet on NDCG@10 and MRR@10 Performance Across Baseline Models (BST, Tri-CDR, FedDCSR, and FedCCTR-LM)}
    \label{fig:Agnostic_Property}
\end{figure*}

To investigate the model-agnostic property of PrivAugNet, we integrated it into three diverse baseline models—BST~\citep{chen2019behavior}, Tri-CDR~\citep{ma2024triple}, and FedDCSR~\citep{zhang2024feddcsr}—and benchmarked their performance against the proposed FedCCTR-LM. As shown in Figure~\ref{fig:Agnostic_Property}, PrivAugNet consistently improves the predictive accuracy of all models across diverse datasets, highlighting its adaptability and robustness.

PrivAugNet incorporates user feature augmentation, item feature enrichment, and interaction sequence expansion, offering flexible enhancements based on the target model’s structure. BST, which fully leverages PrivAugNet’s capabilities, achieves a 5.5\% improvement in NDCG@10 for the Book domain, increasing from 20.97 to 22.13. This result underscores the efficacy of comprehensive augmentation in enhancing sparse features and sequential patterns. Similarly, Tri-CDR, despite utilizing only interaction sequence expansion, demonstrates a substantial 9.1\% improvement in NDCG@10 on the Movie domain. The additional cross-domain signals provided by PrivAugNet effectively mitigate data sparsity, enabling the model to learn more robust representations. FedDCSR, tailored for federated cross-domain CTR prediction, selectively integrates cross-domain interaction sequences while excluding user and item-level augmentations. This approach yields moderate gains, such as a 5.2\% improvement in NDCG@10 on the Food dataset. However, the exclusion of user and item feature enhancements limits its potential.

Despite these notable improvements, none of the PrivAugNet-augmented baseline models outperform FedCCTR-LM. This performance gap is primarily attributable to the unique integration of the IDRA module within FedCCTR-LM, which plays a crucial role in aligning the augmented representations with the original user behavior. By reducing the noise introduced by the LLM-generated augmentations, IDRA ensures that the benefits of PrivAugNet are maximally harnessed, yielding superior performance across all evaluated domains.

\subsection{Visualization of Domain Modeling in FedCCTR-LM (RQ5)}
\label{subsec:exp_5}
\begin{figure*}[!t]
    \centering
    \includegraphics[width=0.8\linewidth, trim={0 0.2cm 0 0}, clip]{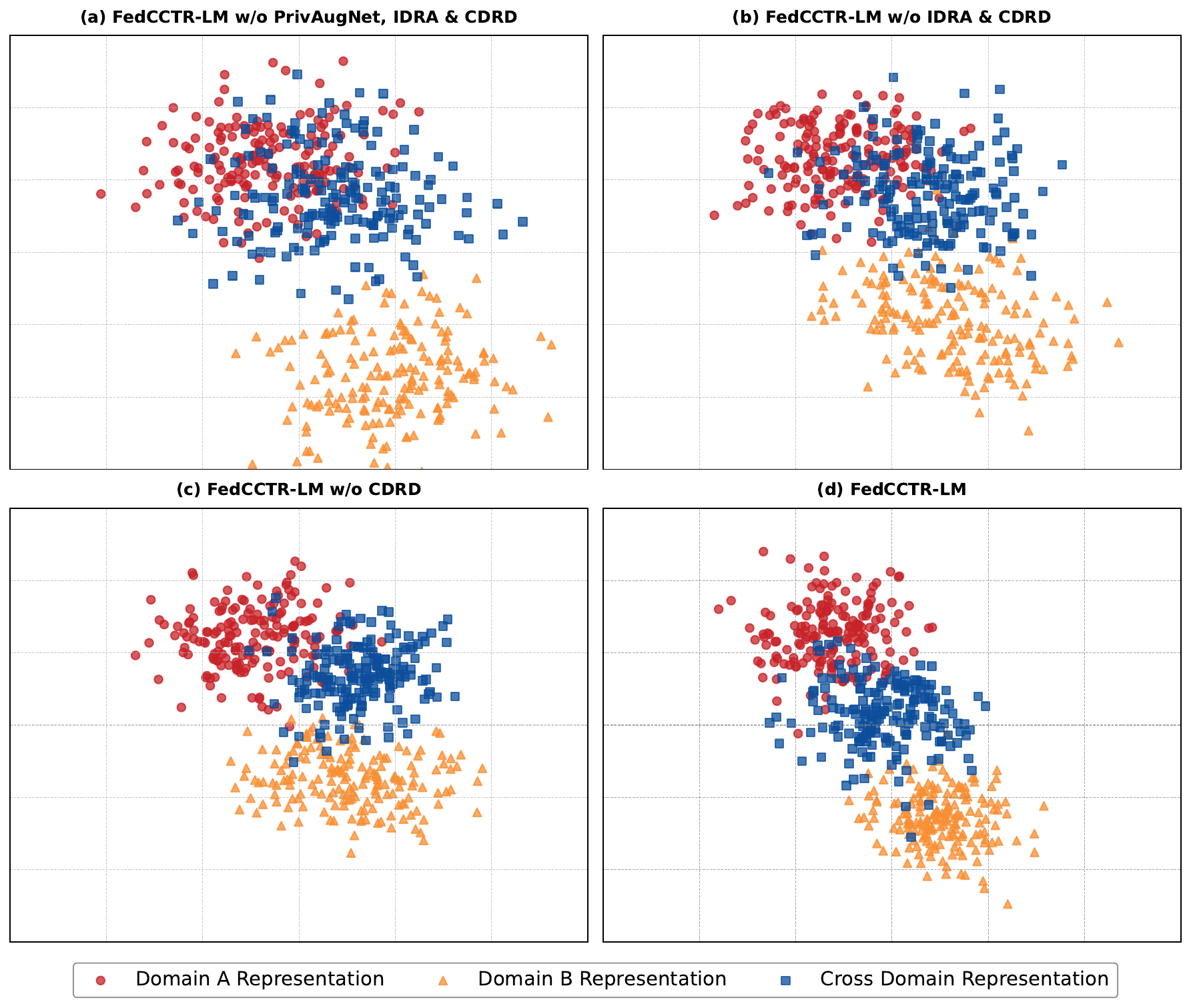}
    \caption{t-SNE Visualization of Domain-specific and Cross-domain Representations in FedCCTR-LM under Varying Ablation Configurations.}
    \label{fig:FedCCTR-LM_Visulization}
\end{figure*}

To visually demonstrate the impact of PrivAugNet, IDRA, and CDRD on domain alignment and disentanglement, we present t-SNE\citep{maaten2008TSNE} visualizations of randomly selected cross-domain sequential representations for 180 users. Figure~\ref{fig:FedCCTR-LM_Visulization} illustrates the distributions of domain-specific representations ($\mathbf{h}^A$ and $\mathbf{h}^B$) and mixed-domain representations ($\mathbf{h}^M$) under four model configurations. Different colors represent and shape distinct domains representation. The results highlight the progressive refinement of domain modeling as each module is incorporated, culminating in robust domain separation and alignment in the full FedCCTR-LM configuration.

\begin{enumerate}
    \item {Figure~\ref{fig:FedCCTR-LM_Visulization} (a):} FedCCTR-LM without PrivAugNet, IDRA, and CDRD struggles to distinguish mixed-domain representations. $\mathbf{h}^M$ aligns closely with $\mathbf{h}^A$, while $\mathbf{h}^B$ is scattered and distant. This lack of coherence suggests poor separation between domains and significant noise in Domain B’s representations, underscoring the critical need for augmentation and alignment mechanisms to improve domain differentiation.

    \item {Figure~\ref{fig:FedCCTR-LM_Visulization} (b):} incorporates PrivAugNet, enhancing feature diversity through user profile enrichment, item augmentation, and sequence expansion. While $\mathbf{h}^B$ begins to shift towards $\mathbf{h}^A$ and $\mathbf{h}^M$, the mixed-domain representation remains too closely aligned with Domain A. Increased dispersion in $\mathbf{h}^A$ and $\mathbf{h}^M$ highlights the trade-off between feature diversity and the introduction of noise, demonstrating the necessity of alignment mechanisms like IDRA to refine augmented features and reduce inconsistency..

    \item {Figure~\ref{fig:FedCCTR-LM_Visulization} (c):} the inclusion of IDRA reduces the dispersion of all domain representations, enhancing intra-domain consistency by mitigating noise from LLM-based augmentations. However, without CDRD, the overlap among $\mathbf{h}^A$, $\mathbf{h}^B$, and $\mathbf{h}^M$ persists, reflecting inadequate separation between shared and domain-specific features. This overlap limits the model’s ability to disentangle cross-domain and intra-domain signals, which is essential for robust generalization.

    \item {Figure~\ref{fig:FedCCTR-LM_Visulization} (d):} demonstrates the full FedCCTR-LM configuration, integrating PrivAugNet, IDRA, and CDRD. Distinct and well-separated clusters for $\mathbf{h}^A$, $\mathbf{h}^B$, and $\mathbf{h}^M$ emerge, with $\mathbf{h}^M$ positioned centrally between the domain-specific clusters. This configuration effectively captures shared user behaviors while preserving domain-specific characteristics. The reduced dispersion and clear cluster boundaries validate the synergy of the three modules in achieving robust alignment, disentanglement, and cross-domain generalization.
\end{enumerate}

\section{Conclusion and Future Study}
\label{sec:conclusion}
This study proposed FedCCTR-LM, an advanced federated framework designed to address three fundamental challenges in CCTR prediction: non-I.I.D. data distributions, cross-domain knowledge transfer, and the privacy-utility trade-off. The framework integrates three key modules to achieve these goals. PrivAugNet employs LLMs to enhance user and item representations while expanding interaction sequences, effectively alleviating data sparsity and feature inconsistency across domains. IDST-CL utilizes independent domain-specific transformers and contrastive learning to disentangle cross-domain preferences while aligning domain-specific representations, enabling nuanced user behavior modeling. AdaLDP dynamically balances privacy preservation and model utility by adapting noise injection to privacy requirements. Comprehensive experiments on real-world datasets demonstrated the superior performance of FedCCTR-LM in achieving significant improvements in CCTR prediction accuracy and privacy compliance over competitive baselines. Visualizations and ablation studies validated the contributions of each module, highlighting the framework’s ability to produce coherent and well-aligned representations across heterogeneous domains.

For future research, we aim to extend this work by exploring multimodal enhancement within the federated framework, leveraging visual and textual content to further enrich the augmented representations provided by PrivAugNet. Additionally, extending FedCCTR-LM to dynamic, real-time federated environments can unlock broader applications in cross-domain recommendation scenarios. These advancements hold the potential to enhance the predictive power and adaptability of federated learning systems while addressing the growing need for robust, privacy-aware AI solutions.






\appendix

\section{Proof of Privacy Guarantees for AdaLDP}
\label{sec:appendix_A}

In this appendix, we provide a detailed proof that the Adaptive Local Differential Privacy (AdaLDP) mechanism satisfies \((\epsilon, \delta)\)-differential privacy in the context of federated learning. We leverage the framework of Rényi Differential Privacy (RDP) and account for the privacy amplification effect due to subsampling.

\subsection{Preliminaries}

\begin{definition}[Rényi Differential Privacy \citep{mironov2017renyi}]
    Let \(D\) and \(D'\) be two neighboring datasets differing in a single element, and let \(\mathcal{M}\) be a randomized algorithm. For any Rényi order \(\alpha > 1\), the mechanism \(\mathcal{M}\) is said to satisfy \((\alpha, \epsilon)\)-Rényi Differential Privacy (RDP) if:
    \begin{equation}
    D_\alpha\left(\mathcal{M}(D) \parallel \mathcal{M}(D')\right) \leq \epsilon,
    \end{equation}
    where the Rényi divergence is defined as:
    \begin{equation}
    D_\alpha(P \parallel Q) = \frac{1}{\alpha - 1} \ln \mathbb{E}_{x \sim Q}\left[ \left( \frac{P(x)}{Q(x)} \right)^\alpha \right].
    \end{equation}
\end{definition}

\begin{lemma}[RDP of the Gaussian Mechanism \citep{mironov2017renyi}]
\label{lemma:gaussian_rdp}
Let \(f\) be a function with \(\ell_2\)-sensitivity \(\Delta\), and define the Gaussian mechanism as:
\begin{equation}
\mathcal{M}(D) = f(D) + \mathcal{N}(0, \sigma^2 \mathbf{I}),
\end{equation}
where \(\mathcal{N}(0, \sigma^2 \mathbf{I})\) denotes the multivariate Gaussian distribution with covariance \(\sigma^2 \mathbf{I}\). Then, for any \(\alpha \geq 1\), the mechanism satisfies \((\alpha, \epsilon)\)-RDP with:
\begin{equation}
\epsilon = \frac{\alpha \Delta^2}{2 \sigma^2}.
\end{equation}
\end{lemma}

\subsection{Gradient Clipping and Sensitivity Analysis}

To bound the sensitivity of gradients, each client \(k\) at round \(t\) clips its local gradient \(\nabla_{k,t}\) using a threshold \(\theta\):
\begin{equation}
\nabla_{k,t}^{\text{clip}} = \nabla_{k,t} \cdot \min\left(1, \frac{\theta}{\| \nabla_{k,t} \|_2}\right).
\end{equation}
This operation ensures that \(\| \nabla_{k,t}^{\text{clip}} \|_2 \leq \theta\). The function \(f\) maps the client's data to the clipped gradient, and its \(\ell_2\)-sensitivity \(\Delta\) is bounded.

\begin{lemma}[Sensitivity of Clipped Gradients]
The function \(f\) has \(\ell_2\)-sensitivity \(\Delta \leq 2\theta\).
\end{lemma}

\begin{pf}
Consider two neighboring datasets \(D\) and \(D'\) differing in one data point. Since each clipped gradient has a norm at most \(\theta\), the maximum change in the gradient norm is:
\begin{equation}
\| \nabla_{k,t}^{\text{clip}}(D) - \nabla_{k,t}^{\text{clip}}(D') \|_2 \leq 2\theta.
\end{equation}
\end{pf}

\subsection{Applying the Gaussian Mechanism}

After clipping, each client adds Gaussian noise to its gradient:
\begin{equation}
\tilde{\nabla}_{k,t} = \nabla_{k,t}^{\text{clip}} + \mathcal{N}(0, \sigma_t^2 \mathbf{I}).
\end{equation}
By Lemma~\ref{lemma:gaussian_rdp}, adding Gaussian noise with variance \(\sigma_t^2\) ensures that the mechanism satisfies \((\alpha, \epsilon_t^{\text{GM}})\)-RDP, where:
\begin{equation}
\epsilon_t^{\text{GM}} = \frac{\alpha \Delta^2}{2 \sigma_t^2} = \frac{\alpha (2\theta)^2}{2 \sigma_t^2} = \frac{2\alpha \theta^2}{\sigma_t^2}.
\end{equation}

\subsection{Privacy Amplification by Subsampling}

In federated learning, only a subset of clients is randomly selected at each round \(t\) with a sampling rate \(\rho\). This introduces a privacy amplification effect. According to the subsampled RDP theorem (Theorem 9 in \citep{wang2019subsampled}), the per-round RDP loss satisfies:

\begin{lemma}[Subsampled Gaussian Mechanism]
For the subsampled Gaussian mechanism with sampling rate \(\rho\), if the single-step RDP loss is \(\epsilon_t^{\text{GM}}\), then the per-round RDP loss \(\epsilon_t(\alpha)\) satisfies:
\begin{equation}
\epsilon_t(\alpha) \leq \frac{1}{\alpha - 1} \ln \left( 1 + \rho^2 \left( e^{ (\alpha - 1) \epsilon_t^{\text{GM}} } - 1 \right) \right).
\end{equation}
\end{lemma}

\begin{pf}
See Theorem 9 in \citep{wang2019subsampled}.
\end{pf}

Substituting \(\epsilon_t^{\text{GM}} = \dfrac{2\alpha \theta^2}{\sigma_t^2}\) into the above, we obtain:
\begin{equation}
\epsilon_t(\alpha) \leq \frac{1}{\alpha - 1} \ln \left( 1 + \rho^2 \left( e^{ \dfrac{2\alpha (\alpha - 1) \theta^2}{\sigma_t^2} } - 1 \right) \right).
\end{equation}

\subsection{Accumulating Privacy Loss}

Since the mechanisms applied at different rounds operate on disjoint data (each client's local data remains the same across rounds), we can sum the per-round RDP losses to obtain the total RDP loss:
\begin{equation}
\epsilon_{\text{total}}(\alpha) = \sum_{t=1}^T \epsilon_t(\alpha).
\end{equation}

\subsection{Converting RDP to \((\epsilon, \delta)\)-Differential Privacy}

To convert the accumulated RDP guarantee to \((\epsilon, \delta)\)-differential privacy, we use the following lemma from \citep{mironov2017renyi}:

\begin{lemma}[Conversion from RDP to \((\epsilon, \delta)\)-DP]
If a mechanism satisfies \((\alpha, \epsilon_{\text{total}}(\alpha))\)-RDP, then for any \(\delta > 0\), it also satisfies \((\epsilon, \delta)\)-differential privacy, where:
\begin{equation}
\epsilon = \epsilon_{\text{total}}(\alpha) - \frac{\ln \delta}{\alpha - 1}.
\end{equation}
\end{lemma}

\begin{pf}
See Proposition 3 in \citep{mironov2017renyi}.
\end{pf}

By selecting an appropriate order \(\alpha\) (e.g., optimizing \(\alpha\) to minimize \(\epsilon\)), we can compute \(\epsilon\) for a given \(\delta\).

\subsection{Privacy Budget Management}

Each client is allocated a total privacy budget \((\epsilon_0, \delta)\). At each round \(t\), the client updates its cumulative RDP loss and checks whether participating in the next round would exceed the privacy budget.

Specifically, the client computes:
\begin{equation}
\epsilon_{\text{cumulative}}(\alpha) = \sum_{i=1}^t \epsilon_i(\alpha).
\end{equation}
Before participating in round \(t+1\), the client estimates the new cumulative RDP loss:
\begin{equation}
\epsilon_{\text{cumulative}}^{\text{new}}(\alpha) = \epsilon_{\text{cumulative}}(\alpha) + \epsilon_{t+1}(\alpha).
\end{equation}
Using the conversion lemma, the client converts \(\epsilon_{\text{cumulative}}^{\text{new}}(\alpha)\) to \((\epsilon, \delta)\)-DP:
\begin{equation}
\epsilon^{\text{new}} = \epsilon_{\text{cumulative}}^{\text{new}}(\alpha) - \frac{\ln \delta}{\alpha - 1}.
\end{equation}
If \(\epsilon^{\text{new}} \leq \epsilon_0\), the client continues to participate in round \(t+1\); otherwise, to ensure that the total privacy loss does not exceed \(\epsilon_0\), the client stops participating.

By carefully calibrating the noise standard deviation \(\sigma_t\) at each round and monitoring the cumulative privacy loss using the RDP framework, AdaLDP ensures that each client maintains their privacy guarantee within the specified budget \((\epsilon_0, \delta)\)-DP.

\printcredits

\section*{Declaration of competing interest}
The authors declare that they have no known competing financial interests or personal relationships that could have appeared to influence the work reported in this paper.

\section*{Data availability}
The datasets used in this paper are public datasets.

\section*{Statement}
During the preparation of this work, the author(s) used ChatGPT 4o from OpenAI for the purpose of English language refinement and enhancing the clarity of technical expressions. After utilizing this tool, the author(s) thoroughly reviewed and edited the content to ensure its accuracy and academic rigor, taking full responsibility for the final content of the published article.

\section*{Acknowledgment}
This research is supported by the National Natural Science Foundation of China (No. 62271274), Natural Science Foundation of Zhejiang Province (No. LZ20F020001), Science and Technology Innovation 2025 Major Project of Ningbo (No. 20211ZDYF020036), Natural Science Foundation of Ningbo (No. 2021J091), the Research and Development of a Digital Infrastructure Cloud Operation and Maintenance Platform Based on 5G and AI (No. HK2022000189), and China Innovation Challenge (Ningbo) Major Project(No. 2023T001).

\bibliographystyle{cas-model2-names}
\bibliography{reference}



\end{document}